\documentclass[twocolumn,twocolappendix]{aastex631}
\usepackage{natbib}
\usepackage{amsmath}
\usepackage{algorithm}
\usepackage{amsfonts}
\usepackage{mathrsfs}
\usepackage{amssymb}
\usepackage{color}
\usepackage{graphicx} 
\let\tablenum\relax
\usepackage{siunitx}
\usepackage{comment}

\newcommand{\XMM}{{XMM-{\it{Newton}}}}
\newcommand{\NuSTAR}{{\it{NuSTAR}}}
\newcommand{\XSPEC}{{\tt{XSPEC}}}

\def \MYTorus {{\tt MYTorus}}

\def \UXClumpy {{\tt \UXClumpy}}
\newcommand{\ms}{\ensuremath{M_{\odot}}}

\begin{document}
%\defcitealias{Zhao2021}{Z21}

\title{An X-ray Significantly Variable, Luminous, Type 2 Quasar at {\it z} = 2.99 with a Massive Host Galaxy}

\author[0000-0002-7791-3671]{Xiurui Zhao}
\affiliation{Department of Astronomy, University of Illinois at Urbana-Champaign, Urbana, IL 61801, USA}
\affiliation{Center for Astrophysics $|$ Harvard \& Smithsonian, 60 Garden Street, Cambridge, MA 02138, USA}

\author[0000-0002-2203-7889]{Stefano Marchesi}
\affiliation{Dipartimento di Fisica e Astronomia (DIFA), Università di Bologna, via Gobetti 93/2, I-40129 Bologna, Italy}
\affiliation{Department of Physics and Astronomy, Clemson University, Kinard Lab of Physics, Clemson, SC 29634, USA}
\affiliation{INAF, Osservatorio di Astrofisica e Scienza dello Spazio di Bologna, via P. Gobetti 93/3, 40129 Bologna, Italy}

\author[0000-0002-6584-1703]{Marco Ajello}
\affiliation{Department of Physics and Astronomy, Clemson University, Kinard Lab of Physics, Clemson, SC 29634, USA}

\author[0000-0002-2115-1137]{Francesca Civano}
\affiliation{NASA Goddard Space Flight Center, Greenbelt, MD 20771, USA}

\author[0000-0001-8121-6177]{Roberto Gilli}
\affiliation{INAF, Osservatorio di Astrofisica e Scienza dello Spazio di Bologna, via P. Gobetti 93/3, 40129 Bologna, Italy}

\author[0000-0001-9094-0984]{Giorgio Lanzuisi}
\affiliation{INAF, Osservatorio di Astrofisica e Scienza dello Spazio di Bologna, via P. Gobetti 93/3, 40129 Bologna, Italy}

\author[0000-0003-4687-8401]{Iván E. López}
\affiliation{INAF, Osservatorio di Astrofisica e Scienza dello Spazio di Bologna, via P. Gobetti 93/3, 40129 Bologna, Italy}

\author[0000-0001-6564-0517]{Ross Silver}
\affiliation{NASA Goddard Space Flight Center, Greenbelt, MD 20771, USA}

\author[0000-0003-3638-8943]{Nuria Torres-Alb\`a}
\affiliation{Department of Physics and Astronomy, Clemson University, Kinard Lab of Physics, Clemson, SC 29634, USA}

\author[0000-0001-9379-4716]{Peter G. Boorman}
\affiliation{Cahill Center for Astrophysics, California Institute of Technology, 1216 East California Boulevard, Pasadena, CA 91125, USA}

\author[0000-0001-6412-2312]{Andrealuna Pizzetti}
\affiliation{Department of Physics and Astronomy, Clemson University, Kinard Lab of Physics, Clemson, SC 29634, USA}

\begin{abstract}
We present a comprehensive X-ray analysis and spectral energy distribution (SED) fitting of WISEA J171419.96+602724.6, an extremely luminous type 2 quasar at $z$ = 2.99. The source was suggested as a candidate Compton-thick (column density N$_{\rm H}>$1.5 $\times$ 10$^{24}$~cm$^{-2}$) quasar by a short \XMM\ observation in 2011. We recently observed the source with deep \NuSTAR\ and \XMM\ exposures in 2021 and found that the source has a lower obscuration of N$_{\rm H}\sim$5 $\times$ 10$^{22}$~cm$^{-2}$ with an about four times lower flux. The two epochs of observations suggested that the source was significantly variable in X-ray obscuration, flux, and intrinsic luminosity at 2--3~$\sigma$ in less than 2.5 years (in the source rest frame). We performed SED fitting of this source using CIGALE thanks to its great availability of multiwavelength data (from hard X-rays to radio). The source is very luminous with a bolometric luminosity of $L_{\rm BOL}\sim$ 2.5 $\times$ 10$^{47}$~erg~s$^{-1}$. Its host galaxy has a huge star formation rate (SFR) of $\sim$1280 M$_\sun$~yr$^{-1}$ and a huge stellar mass of $\sim$1.1 $\times$ 10$^{12}$ M$_\sun$. The correlation between the SFR and stellar mass of this source is consistent with what was measured in the high-$z$ quasars. It is also consistent with what was measured in the main-sequence star-forming galaxies, suggesting that the presence of the active nucleus in our target does not enhance or suppress the SFR of its host galaxy.
The source is an Infrared hyper-luminous, obscured galaxy with significant amount of hot dust in its torus and shares many similar properties with hot, dust obscured galaxies.
\end{abstract}

\section{Introduction} \label{sec:intro}

It is widely believed that the powerful emission of active galactic nuclei (AGN) originates from mass accreting onto the central supermassive black hole \citep[SMBH,][]{Salpeter1964}. Thus, AGN activity can be used as a tracer of SMBH growth. These black holes (BHs) are found to have a mass ranging from 10$^6$ to 10$^{10}$\ms\ \citep{Mortlock2011}.  %At the high end, massive, $\ge$10$^9$ \ms, BHs detected at high-$z$ \citep{Banados2018} require continuous near Eddington-limit accretion of the gas surrounding the SMBH. 
The large amount of material needed for quick BH growth, particularly near the peak of BH accretion history $z\sim$1.5--3, is expected to obscure the nuclear emission, implying that SMBH growth may be obscured \citep{Menci2008,Merloni2008,Delvecchio2014}. 

Simulations of cosmic X-ray background (CXB) show that the majority of quickly growing SMBHs in this redshift range are obscured by material with gas column density $\ge$10$^{23}$\,cm$^{-2}$, and $\sim$40\% of them are obscured by Compton-thick (CT-; N$\rm _H\ge$ 10$^{24}$\,cm$^{-2}$) gas \citep{gilli07,Buchner2015}. This rapid BH growth phase, although extremely important for the BH cosmic accretion history, is mostly inaccessible due to the lack of detections of luminous, high-$z$, obscured AGN. Selection techniques based on mid-IR emission have been developed to select candidate CT-AGN up to $z$ = 2--3 \citep[e.g.,][]{Daddi2007}. However, the measured obscuring column densities are largely uncertain due to the lack of (high signal-to-noise ratio, S/N) X-ray spectra. Currently, most of the X-ray selected AGN at high-$z$ are less obscured, given the bias against detecting obscured (and thus fainter) AGN. Indeed, only a handful of CT-AGN (candidates) have been observed in X-rays at $z\gtrsim1.5$ \citep[e.g.,][]{Gilli2011,Lanzuisi2018,Snios2020,Vito2020}. 

WISEA J171419.96+602724.6 (hereafter, J1714+6027 ) is a type 2 quasar\footnote{Here we define AGN as quasar with intrinsic 2--10~keV luminosity $L_{2-10}\ge10^{44}$~erg~s$^{-1}$.} at $z$ = 2.99. It was selected from the 4XMM-DR10 catalog \citep[4XMM J171419.3+602721,][]{Webb2020} as a CT-quasar candidate. The catalog reports that the source has a hardness ratio (HR) of HR = 0.90$_{-0.16}^{+0.10}$, suggesting that J1714+6027 is a heavily obscured quasar. HR is defined as (H-S)/(H+S), where H and S are hard (2--4.5~keV) and soft (1--2~keV) fluxes. The source was observed with the \XMM\ MOS2 camera in March 2011 (ObsID: 0651370901), at an off-axis angle of 14\farcm8. The exposure of the observation is 11.6~ks, but the vignetting-corrected exposure at the source position is only 2.6~ks, resulting in a total number of 12 net counts. %We extracted the source spectrum following the standard method as detailed in Section.~\ref{sec:XMM}. 
We fitted the source spectrum with a simplified phenomenological model (see more details in Section~\ref{sec:phe}) assuming a photon index of $\Gamma$ = 1.80 \citep[a typical value for type 2 AGN, e.g.,][]{Ricci2017}. The best-fit `line-of-sight' column density is N$\rm _{H,Z}$ = 1.1$_{-0.6}^{+1.9}$ $\times$ $10^{24}$\,cm$^{-2}$, suggesting that the source is a CT-quasar candidate. The source flux is $\sim$1.6 $\times$ 10$^{-13}$ erg~s$^{-1}$ in the 2--10~keV band, implying that the source is bright enough for deep X-ray observations follow-up. Therefore, we proposed the source to \NuSTAR\ and \XMM\ to confirm its CT nature. 

The paper is structured as follows. In Section~\ref{sec:data}, we present the data reduction and spectra analysis of the new X-ray data of the source. In Section~\ref{sec:multi}, we present the multiwavelength data of the source and perform the spectral energy distribution (SED) fitting. In Section~\ref{sec:discuss}, we discuss the X-ray and multiwavelength properties (including its variability properties) of the AGN and the properties of its host galaxy. 

Uncertainties are quoted at the 90\% confidence level throughout the paper unless otherwise stated. The magnitudes used here are in the AB system. Standard cosmological parameters are adopted as follows: $H_0 = 70$~km s$^{-1}$ Mpc$^{-1}$, $\Omega_M = 0.30$, and $\Omega_\Lambda = 0.70$. 
%%%%%%%%%%%%%%%%%%

\begingroup
\renewcommand*{\arraystretch}{1.1}
\begin{table*}
\begin{center}
\caption{Summary of \NuSTAR\ and \XMM\ observations.}\label{Tab:obs}
  \begin{tabular}{cccccc}
       \hline
       \hline
    Instrument&Sequence&Start Time&End Time&Exposure&Net Count Rate\tablenotemark{a}\\ 
    &ObsID&(UTC)& (UTC)&(ks)&$10^{-3}$counts s$^{-1}$\\
    \hline
    \NuSTAR&60701056002&2021-09-27 T22:36:09&2021-09-30 T05:11:09&108/107& 0.82/0.73 \\
    \XMM&0890450101&2021-09-28 T06:05:55&2021-09-28 T23:51:51&30/36/17\tablenotemark{b}&1.93/2.42/8.11\\
       \hline
\end{tabular}
\par
\vspace{.3cm}
\end{center}
\tablenotemark{a}{\footnotesize { The reported \NuSTAR\ count rates are those of the FPMA and FPMB modules in the 3--16\,keV range, respectively. The reported \XMM\ count rates are those of the MOS1, MOS2, and pn modules in the 0.5--10\,keV range, respectively.}}

\tablenotemark{b}{\footnotesize {The reported \XMM\ exposures are the cleaned exposure times for MOS1, MOS2, and pn, respectively after removing the high-background intervals. The on-source exposure is originally 68~ks.\\}}
\end{table*}
\endgroup

\section{X-ray Data Analysis} \label{sec:data}
%\subsection{Source Selection}
J1714+6027 was observed in \NuSTAR\ cycle 7 (PI: Zhao, ID: 7250) on September 27th, 2021 with 108~ks \NuSTAR\ and 68~ks \XMM\ exposures. The details of the observations can be found in Table~\ref{Tab:obs}. We reduce the \NuSTAR\ and \XMM\ data in Section~\ref{sec:reduction} and analyze the spectra in Section~\ref{sec:spec_ana}. We also re-analyze the 2011 \XMM\ observation.

\subsection{Data Reduction} \label{sec:reduction}
Both \NuSTAR\ and \XMM\ observations were reduced using the most updated version of calibration files and software as detailed in the following sections.
\subsubsection{\NuSTAR}
The \NuSTAR\ data were processed using HEASoft v.6.32.1 and \NuSTAR\ Data Analysis Software (NuSTARDAS) v.2.1.2 with the updated calibration and response files CALDB v.20231017. The level 1 raw data were calibrated, cleaned, and screened by running the \texttt{nupipeline} tool. As the background event rates are slightly elevated around the SAA, we used the parameters \texttt{saamode=OPTIMIZED} and \texttt{tentacle=yes} when reducing the data. The sources spectra ancillary response files (ARF) and response matrix files (RMF) are obtained using the \texttt{nuproducts} script. The source spectra are extracted from a 40$^{\prime\prime}$ circular region, corresponding to an encircled energy fraction (EEF) of 60\% at 10~keV. The spectral extraction radius was determined following the method in \citet{Zappacosta_2018} which maximizes both the S/N and the number of net counts. We note that two X-ray sources WISEA J171430.62+602722.7 (J171430+602722, hereafter) and WISEA J171430.09+602635.7 (J171430+602635, hereafter) are about 80\arcsec\ and 90\arcsec\ from J1714+6027. They are 54\% and 70\% fainter than J1714+6027 in the 2--10~keV band, respectively. We also notice another X-ray source, WISEA J171425.30+602928.1 (J171425+602928, hereafter) that is about 125\arcsec\ from J1714+6027 but is 180\% brighter than J1714+6027 (in the 2--10~keV band). The selected source spectra extraction region thus includes $\sim$3\% of the fluxes from the first two sources and includes about 0.8\% of the flux from the third source in the 2--10~keV band. The details of the spectral analysis and source properties of the three sources are described in \ref{sec:near_source}. Therefore, a limited fraction of about 4.5\% of the J1714+6027 flux measured by \NuSTAR\ in the 2--10~keV band is likely to originate from the other three nearby sources. At 10--16~keV, the contaminating flux is about 2.8\% based on the best-fit models of the three nearby sources and J1714+6027.

As the number of counts from the background is comparable to those from the source and \NuSTAR\ background is highly spatially uneven across the field-of-view (FoV), we analyzed the \NuSTAR\ background using the {\tt nuskybgd} tool\footnote{\url{https://github.com/NuSTAR/nuskybgd}} \citep{Wik_2014}. This tool provides more accurate modeling of the \NuSTAR\ background, and it has been widely used in the studies of extended or faint targets and extragalactic surveys. We followed the standard method when applying {\tt nuskybgd} to our data \citep[see, e.g.,][]{Wik_2014,Zhao2021}. The accuracy of the simulated background was tested using the same method as in \citet{Zhao2021} before it was used in the spectral analysis. We note that the source flux measured in \NuSTAR\ is consistent with that measured in \XMM\ (see Section~\ref{sec:spec_ana}), indicating that the \NuSTAR\ background is well simulated. Both FPMA and FPMB spectra are dominated by the background at $>$16~keV, therefore, we only analyzed the \NuSTAR\ data between 3 and 16~keV.

\begingroup
\renewcommand*{\arraystretch}{1.1}
\begin{table*}
\centering
\caption{Best-fit results of the 2011 \XMM\ spectrum and 2021 \NuSTAR+\XMM\ spectra with different models. }
\label{Table:best-fit}
  \begin{tabular}{l|cc|cccc}
       \hline
       \hline       
       &Phenome&Phenome&Phenome&MYT$_{90}$&MYT$_{0}$&borus\\
       \hline
       Year&2011&2011&2021&2021&2021&2021\\
       \hline
       $C$/d.o.f.&10/7&10/7&41/52& 41/52& 40/52& 40/52\\
       $\Gamma$\footnote{The photon index ($\Gamma$) of the {\tt MYTorus} and {\tt borus} models is only allowed to vary between 1.4 to 2.6. $u$ means that the 90\% confidence lower limit of the photon index is not measured at $\Gamma\ge$1.4.}&1.80$^{fix}$&1.45$^{fix}$&1.45$^{+0.19}_{-0.18}$&{1.54$^{+0.20}_{-u}$}&{1.55$^{+0.19}_{-u}$}&1.50$^{+0.18}_{-u}$\\
       $\theta\rm _{obs}$\footnote{Angle between the torus axis and the ``line-of-sight'' direction in degree, e.g., $\theta\rm _{obs}$ = 90$^\circ$ is edge-on and $\theta\rm _{obs}$ = 0$^\circ$ is face-on.}&...&...&...&90$^{fix}$&0$^{fix}$&{60$^{fix}$}\\
       A$_S$&...&...&...&1$^{fix}$&1$^{fix}$&...\\
       N$\rm _{H,Z}$\footnote{``line-of-sight'' column density in $10^{22}$\,cm$^{-2}$.}&114$_{-64}^{+188}$&95$_{-59}^{+172}$&5.9$_{-3.6}^{+4.5}$&{6.3$_{-3.7}^{+4.6}$}&{5.8$_{-3.6}^{+4.4}$}&{5.6$_{-3.1}^{+4.5}$}\\
       N$\rm _{H,S}$\footnote{``global average'' column density of the torus in $10^{24}$\,cm$^{-2}$.}&...&...&...&{1.4$^{fix}$}&{1.4$^{fix}$}&{1.4$^{fix}$}\\
       $c\rm _{f}$\footnote{Covering factor of the torus.}&...&...&...&...&...&{0.20$^{fix}$}\\
       F$_{0.5-2}$\footnote{0.5--2\,keV flux in $10^{-14}$\,erg\,cm$^{-2}$ s$^{-1}$.}&0.5$_{-0.5}^{+0.7}$&0.5$_{-0.5}^{+0.7}$&1.1$_{-0.1}^{+0.1}$&1.1$_{-0.2}^{+0.2}$&1.1$_{-0.1}^{+0.2}$&1.1$_{-0.1}^{+0.2}$\\
       F$_{2-10}$\footnote{2--10\,keV flux in $10^{-14}$\,erg\,cm$^{-2}$ s$^{-1}$.}&16$_{-9}^{+13}$&17$_{-9}^{+15}$&4.6$_{-0.6}^{+0.6}$&4.6$_{-0.6}^{+0.5}$&4.6$_{-0.6}^{+0.6}$&4.6$_{-0.6}^{+0.6}$\\
%       L$_{0.5-2}$\footnote{0.5--2\,keV intrinsic luminosity in $10^{45}$\,erg\,s$^{-1}$.}&9.0&3.2&0.7$_{-0.5}^{+0.5}$&0.5$_{-u}^{+u}$&0.6&0.7\\
       L$_{2-10}$\footnote{2--10\,keV intrinsic luminosity in $10^{45}$\,erg\,s$^{-1}$. The luminosity is derived using `clum' in {\tt xspec}.}&14$_{-8}^{+16}$&9$_{-5}^{+9}$&1.8$_{-0.5}^{+0.5}$&{1.8$_{-0.4}^{+0.7}$}&{ 1.7$_{-0.5}^{+0.6}$}&{ 1.7$_{-0.3}^{+0.7}$}\\
       L$_{10-40}$\footnote{10--40\,keV intrinsic luminosity in $10^{45}$\,erg\,s$^{-1}$.}&16$_{-10}^{+19}$&17$_{-10}^{+18}$&3.5$_{-0.6}^{+0.5}$&{ 3.2$_{-0.5}^{+0.5}$}&{2.9$_{-0.4}^{+0.5}$}&{3.1$_{-0.4}^{+0.5}$}\\
       \hline
	\hline
\end{tabular}
\end{table*}
\endgroup

\subsubsection{\XMM}\label{sec:XMM}
The \XMM\ observation was taken quasi-simultaneously to the \NuSTAR\ one with the EPIC CCD cameras \citep[pn;][]{pn} and two MOS cameras \citep[][]{MOS}: the \XMM\ observation started at the same time, but ended $\sim$9 hours before the \NuSTAR\ one. We reduced the \XMM\ data using the Science Analysis System \citep[SAS;][]{SAS} version 16.1.0. The light curves of the three cameras are produced using the \texttt{evselect} tool at $>$10~keV with \texttt{PATTERN==0} (where the events are dominated by the backgrounds rather than the sources in the field-of-view, FoV). Due to the strong background flares (which might be associated with a C-level solar flare\footnote{\url{https://www.swpc.noaa.gov/products/goes-x-ray-flux}} during the \XMM\ observation), 47--75\% of the \XMM\ exposure was lost, where we removed the time interval with count rates in the FoV larger than 0.15, 0.20, and 0.40~cts/s in MOS1, MOS2, and pn, respectively. Therefore, the effective exposures used for spectral analysis of the three cameras are 30, 36, and 17~ks, respectively. The source spectra are extracted from a 15\arcsec\ circular region, corresponding to $\approx$70\% of the EEF at 1.5\,keV, while the background spectra are obtained from a 40\arcsec\ circle located near the source. We visually inspected the \XMM\ image to avoid contamination to the background from sources near J1714+6027. 

We analyzed the \XMM\ spectra between 0.5 and 10~keV in all three cameras. The net count rates of the two modules of \NuSTAR\ and three cameras of \XMM\ are reported in Table~\ref{Tab:obs}. The \NuSTAR\ and \XMM\ spectra are grouped with a minimum of 30 and 10 counts per bin using \texttt{grppha}, respectively.

\subsection{Spectral Analysis} \label{sec:spec_ana}
We performed a broadband (0.5--16~keV) spectral analysis of J1714+6027 using \texttt{XSPEC} \citep{Arnaud1996} version 12.13.1. The photoelectric cross-section is from \cite{Verner1996}; the element abundance is from \citet{Anders1989} and metal abundance is fixed to solar; the Galactic absorption column density is fixed at 2.2 $\times$ 10$^{20}$ cm$^{-2}$ \citep[\texttt{nh} task,][]{nh}. The C statistic \citep{Cash1979} is adopted when fitting the spectra. The source redshift is fixed at $z$ = 2.99 \citep{Lacy2007}.

We first analyzed the spectra using a simplified phenomenological model considering only the absorption along the line-of-sight. As the spectra of the heavily obscured AGN were not found to be well described by a simple line-of-sight component, especially at the Compton hump at 10--30~keV \citep[e.g.,][]{Lightman1988,pexrav,MYTorus2012}, we then fit the spectra of J1714+6027 using two physically-motivated models {\tt MYTorus}  { \citep{MYTorus2009} and {\tt borus} \citep{Borus}} including also the reflection from the absorber, which was widely used to model the spectra of heavily obscured AGN with high-quality data { \citep[e.g., ][]{puccetti14,Annuar15,Marchesi2018,Ursini18, Nuria}}. The best-fit results are reported in Table~\ref{Table:best-fit}.

\subsubsection{Simplified Phenomenological Model}\label{sec:phe}
We first fit the source spectra with a simplified phenomenological absorbed power-law model as described in equation~\ref{eq:powerlaw}. The {\tt constant} models the cross-calibration between \NuSTAR\ and \XMM, noted as $C_{NuS/XMM}$. {\tt phabs} models the Galactic absorption. {\tt zphabs} models the quasar intrinsic absorption along the line-of-sight caused by the obscuring gas and dust surrounding the accreting supermassive black hole. {\tt zpowerlw} models the quasar intrinsic X-ray emission produced by the hot corona. The phenomenological model (Model A), in \XSPEC\ terminology, is thus:
\begin{equation}\label{eq:powerlaw}
\begin{aligned}
Model A = &constant*phabs*(zphabs*zpowerlw)
\end{aligned}
\end{equation}

\begin{figure*} 
\begin{minipage}[b]{.49\textwidth}
\centering
\includegraphics[width=\textwidth]{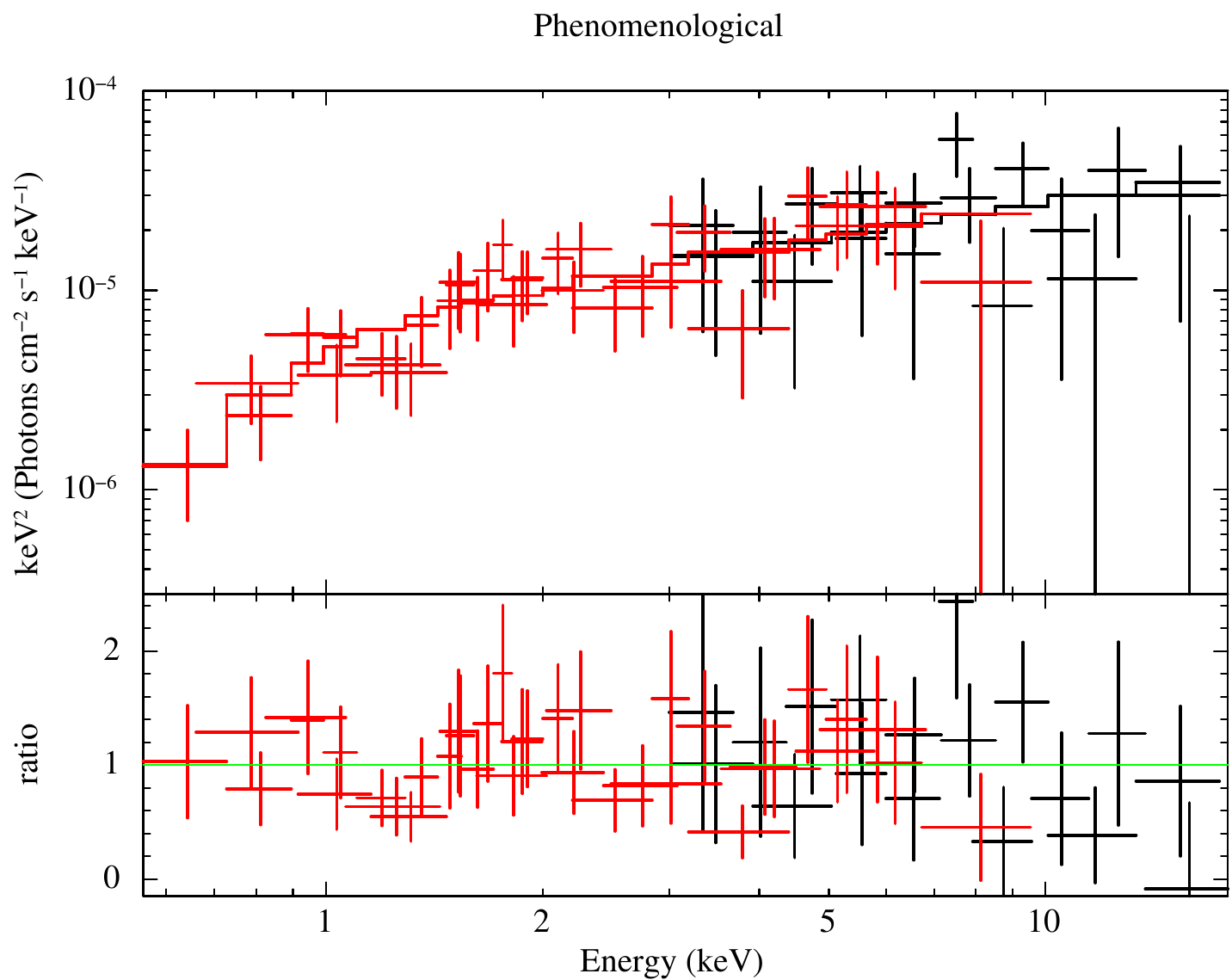}
\includegraphics[width=\textwidth]{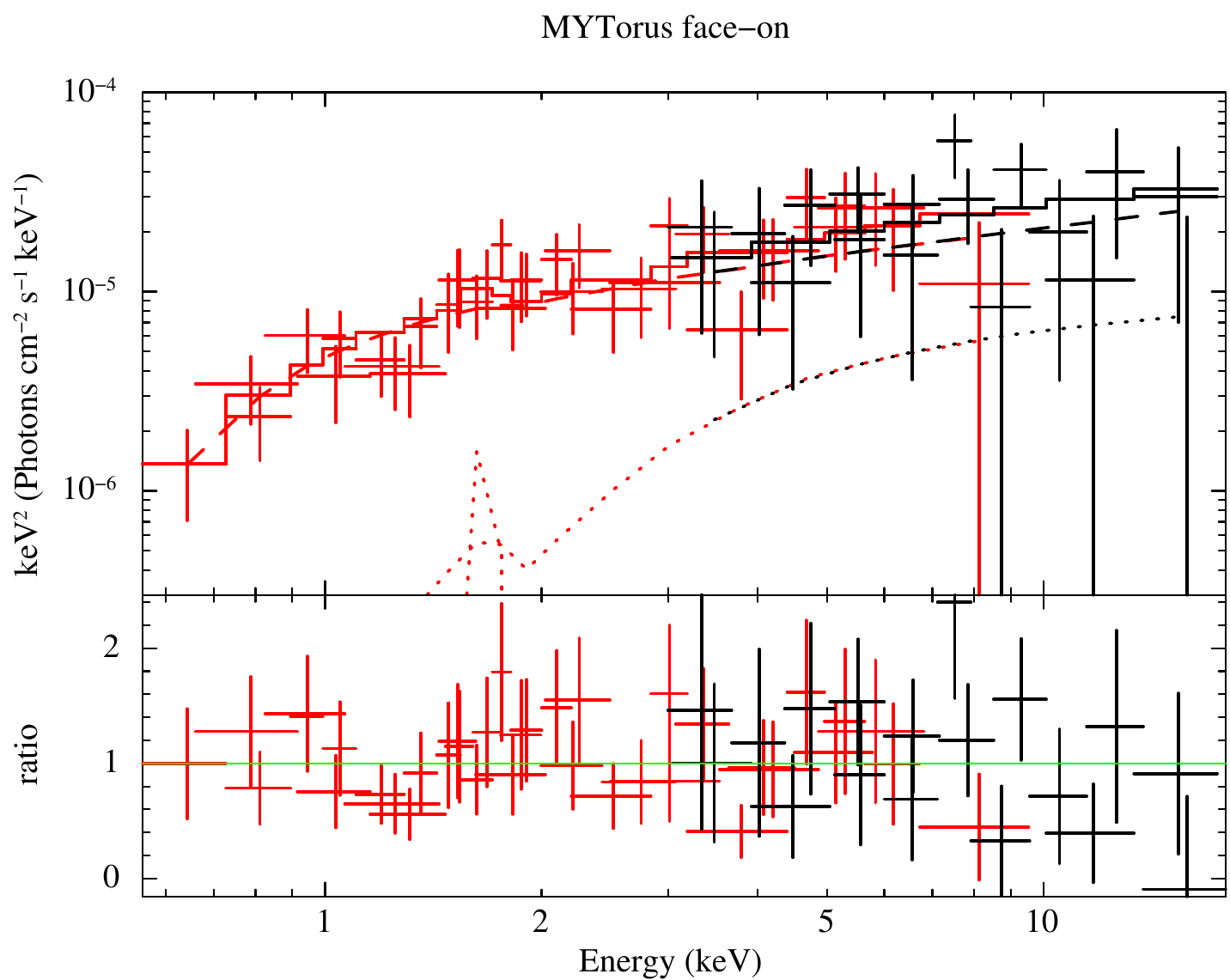}
\end{minipage}
\begin{minipage}[b]{.49\textwidth}
\centering
\includegraphics[width=\textwidth]{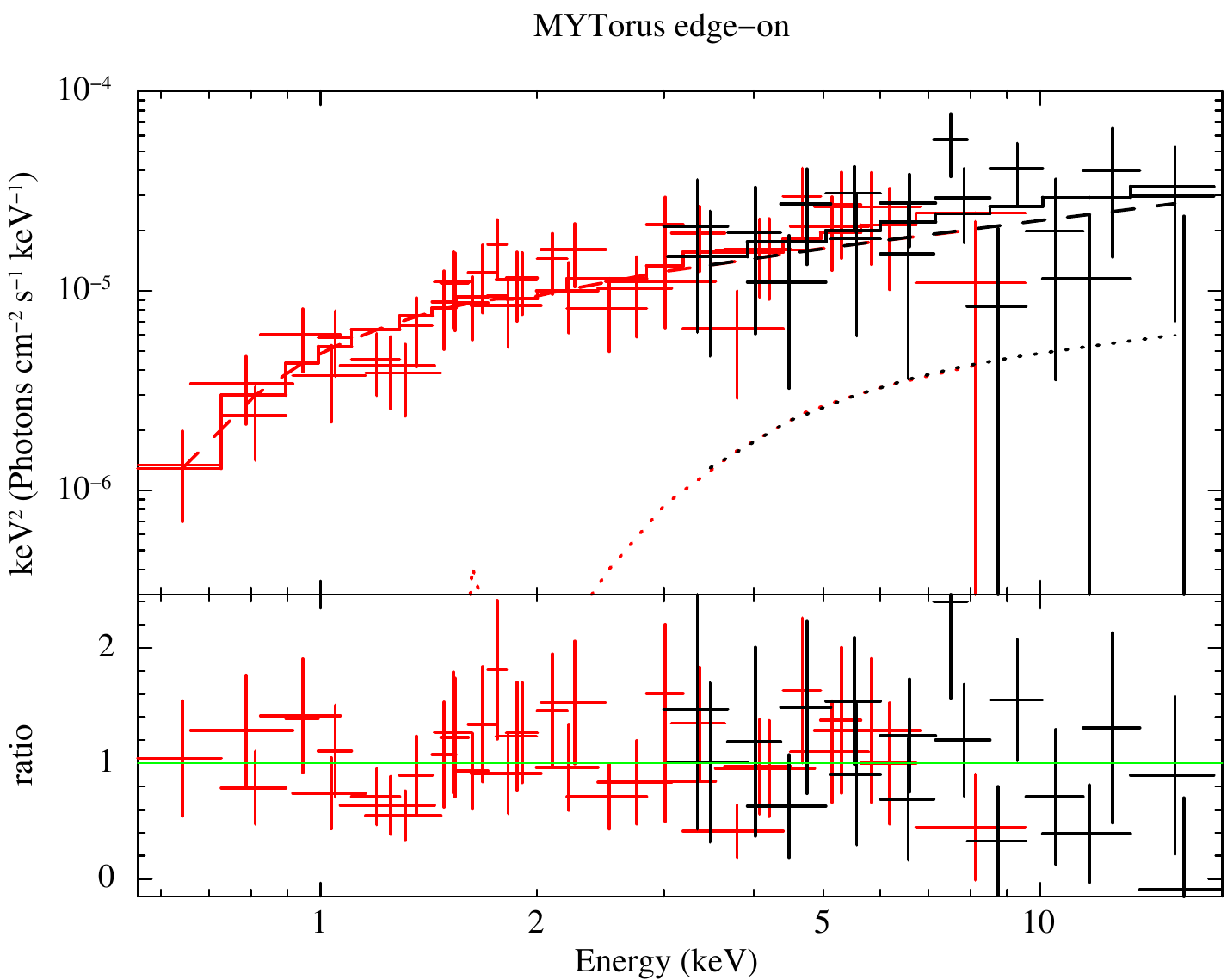}
\includegraphics[width=\textwidth]{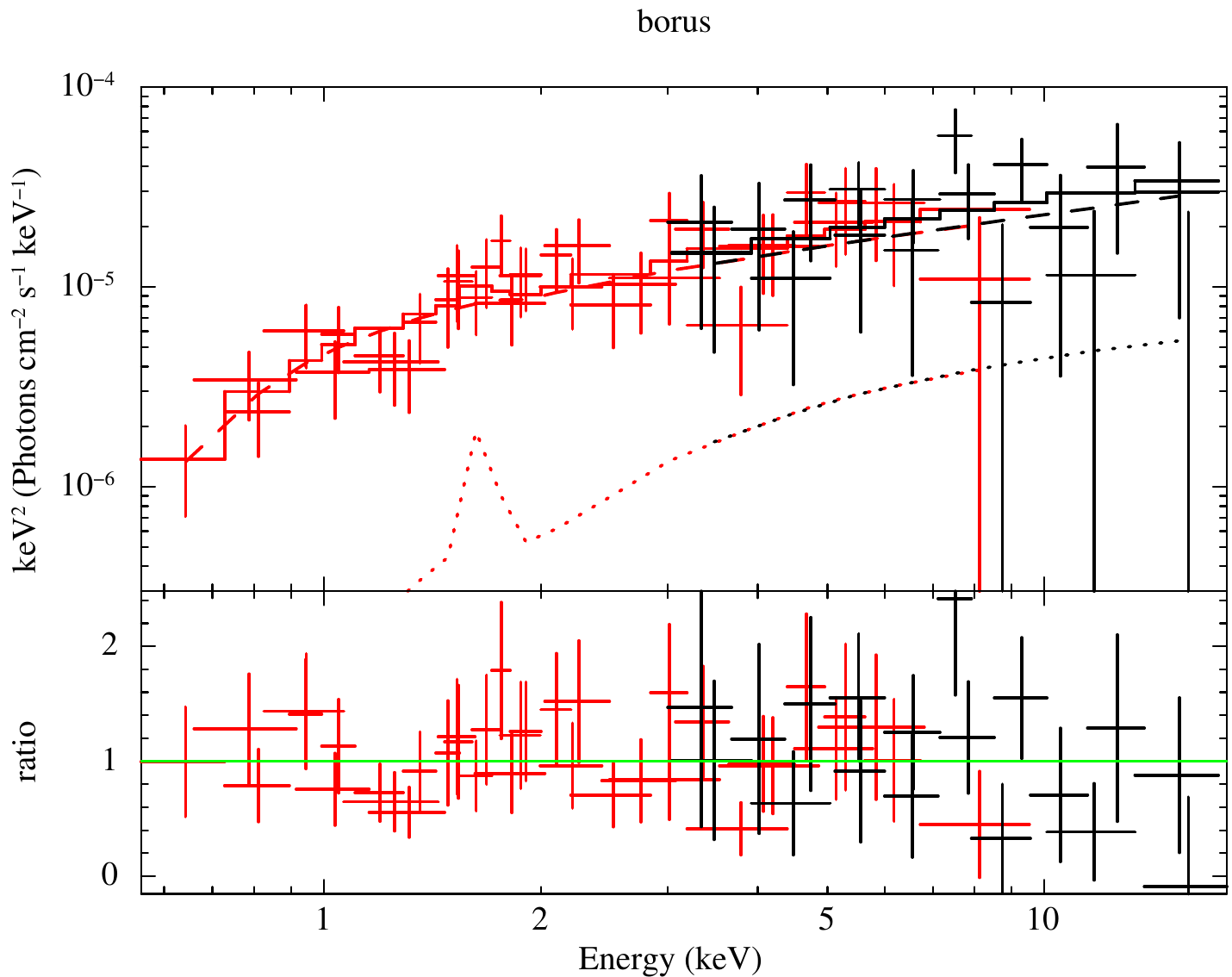}
\end{minipage}
\caption{X-ray spectra (in the observed frame) of J1714+6027 fitted using phenomenological, \MYTorus\ (edge-on and face-on), and \texttt{borus} models. The black and red points correspond to \NuSTAR\ and \XMM\ data. The solid, dashed, and dotted lines correspond to the model predictions from the total, line-of-sight component, and reflection components.}
\label{fig:spectrum}
\end{figure*}

We first leave the $C_{NuS/XMM}$ free to vary when fitting the spectra, which allows us to check the background simulation of \NuSTAR. The best-fit cross-calibration is $C_{NuS/XMM}$ = 1.08$_{-0.20}^{+0.49}$, which is consistent with the cross-calibration between \NuSTAR\ and \XMM\ determined by the International Astronomical Consortium for High Energy Calibration \citep[IACHEC,][]{Madsen2017}. Due to the limited number of photons, we fixed $C_{NuS/XMM}$ at unity when fitting the spectra to better constrain the other source properties. The best-fit photon index is $\Gamma$ = 1.45$^{+0.19}_{-0.18}$, which is harder than the typical photon indices of an obscured quasar with $\Gamma\sim$1.80 \citep[e.g.,][]{Ricci2017}. The best-fit line-of-sight obscuration is N$\rm _{H,Z}$ = 5.9$_{-3.6}^{+4.5}$ $\times$ $10^{22}$\,cm$^{-2}$, suggesting that the source is an obscured Compton-thin ($10^{22}$\,cm$^{-2}$ $<$ N$\rm _{H}<$ $10^{24}$\,cm$^{-2}$) quasar. The best-fit results of the phenomenological model are listed in Table~\ref{Table:best-fit} and the source spectra with best-fit models are plotted in Fig.~\ref{fig:spectrum}. We notice that the overall spectral shape is well-fitted. The C-statistics divided by degrees of freedom (d.o.f., hereafter) is $C$/d.o.f = 41/52.

\subsubsection{Physical Model: {\tt MYTorus}}
We then fit the spectra with a physically-motivated model {\tt MYTorus} \citep{MYTorus2009}, which models both the line-of-sight continuum of the quasar and the intrinsic emission reflected from the torus. The basic geometry of the {\tt MYTorus} model consists of a torus that has a fixed half-opening angle, $\theta_{\rm oa}$ = 60$^{\circ}$ or a fixed covering factor of $f_c$ = 0.5, with a circular cross-section. The line-of-sight continuum is modeled by the absorption ({\tt MYTZ}) on a power law. The reflection component is modeled by a Compton-scattered continuum ({\tt MYTS}) and the most prominent fluorescent emission lines ({\tt MYTL}), i.e., the Fe K$\alpha$ and Fe K$\beta$ lines, at 6.4\,keV and 7.06\,keV, respectively. It is worth noting that if the geometry of the torus differs significantly from this one, or if there is a non-negligible time delay between the intrinsic continuum emission and the Compton-scattered continuum one, i.e., the central region is not compact and the intrinsic emission varies rapidly, the scattered component normalization can significantly differ from the main component one. To take these effects into account, the scattered continuum and the emission line component are multiplied by a constant, which we define here as $A_S$. 

The configurations of {\tt MYTorus} are very flexible. We used the `decoupled' configuration of {\tt MYTorus} as introduced in \citet{MYTorus2012}, assuming that the line-of-sight continuum and the reflection component can in principle have different inclination angles ($\theta_{\rm obs}$, the angle between the torus axis and the line-of-sight) and column density values. In the `decoupled' configuration, the line-of-sight absorption ({\tt MYTZ}) is modeled with $\theta_{\rm obs,Z}$ = 90$^\circ$, so that the derived N$\rm _{H,Z}$ is the line-of-sight column density rather than the column density along the equator \citep[see, e.g., ][for more details]{Zhao_2019_1}. The inclination of J1714+6027 can be between 0$^\circ$ and 90$^\circ$, which can be modeled by using two sets of reflection components and using two different $A_S$. Here we fit the spectra assuming an `edge-on' scenario ($\theta_{\rm obs,S}$ = 90$^\circ$) and a `face-on' scenario ($\theta_{\rm obs,S}$ = 0$^\circ$). However, we note that a much better spectral quality is needed to achieve meaningful constraints on the source properties. 

In \XSPEC\ the \MYTorus\ model is described as follows:
\begin{equation}\label{eq:coupled}
\begin{aligned}
Model~B =&constant*phabs*(MYTZ*zpowerlw\\
&+A_S*MYTS+A_S*MYTL)
\end{aligned}
\end{equation}

{ For the `edge-on' scenario, the best-fit photon index is $\Gamma$ = 1.54$^{+0.20}_{-u}$ ({\tt MYTorus} and {\tt borus} allows the photon index to vary between 1.4 and 2.6). The best-fit line-of-sight column density is N$\rm _{H,Z}$ = 6.3$_{-3.7}^{+4.6}$ $\times$ $10^{22}$\,cm$^{-2}$. The torus column density N$\rm _{H,S}$ and $A_S$ are entirely unconstrained. Therefore, we fixed $A_S$ at $A_S$ = 1 and fixed N$\rm _{H,S}$ at the average torus column density N$\rm _{H,S,Zhao2021}\sim$1.4 $\times$ $10^{24}$\,cm$^{-2}$ derived from a sample of $\sim$100 low--redshift Compton-thin AGN \citep{Zhao2021a}.

For the `face-on' scenario, the best-fit photon index is $\Gamma$ = 1.55$^{+0.19}_{-u}$. The best-fit line-of-sight column density is N$\rm _{H,Z}$ = 5.8$_{-3.6}^{+4.4}$ $\times$ $10^{22}$\,cm$^{-2}$. We fixed $A_S$ at $A_S$ = 1 and fixed N$\rm _{H,S}$ at N$\rm _{H,S}$ = 1.4 $\times$ $10^{24}$\,cm$^{-2}$ as well as they are entirely unconstrained. }

\subsubsection{Physical Model: {\tt Borus}}
We now fit the spectra using {\tt borus} \citep{Borus}, another physically-motivated model, which can be used to compare the results with those obtained from the {\tt MYTorus} model. {\tt borus} assumes the geometry of the torus as a sphere with conical cutouts at both poles, so it allows for variable covering factors. The {\tt borus} model is composed of a reprocessed component that is reflected by a toroidal structure from the intrinsic emission. Besides the reflection component, we added a `line-of-sight' component modeled by an absorbed cutoff power-law. In XSPEC, the {\tt borus} model is described as:
\begin{equation}\label{eq:Borus}
Model C =constant*phabs*(borus+zphabs*cabs*cutoffpl) 
\end{equation}

where $zphabs$*$cabs$ models the absorption along the `line-of-sight', which includes the effect of Compton scattering. The cutoff energy $E_{\rm cut}$ of the power law was found to be unconstrained due to the low quality of the data, so we fix $E_{\rm cut}$ at 150~keV as found in sources with similar luminosities \citep{Lanzuisi2019} to reduce the degeneracy. {The best-fit photon index is $\Gamma$ = 1.50$^{+0.18}_{-u}$. The best-fit line-of-sight column density is N$\rm _{H,Z}$ = 5.6$_{-3.1}^{+4.5}$ $\times$ $10^{22}$\,cm$^{-2}$. The best-fit inclination (the angle between the line-of-sight and the torus axis) favors a face-on scenario but is not constrained ($\theta$ = 18$^{+55}_{-u}$ deg at 1~$\sigma$ confidence level). To reduce the degeneracy, we fixed the inclination angle at $\theta$ = 60$^{\circ}$ as suggested in \citet{Zhao_2020}. The torus column density and torus covering factor are both unconstrained. Therefore, we fixed the torus column density at N$\rm _{H,S}$ = 1.4 $\times$ $10^{24}$\,cm$^{-2}$. We fixed the torus covering factor at $c\rm _{f}$ = 0.2, which was measured in sources with X-ray luminosity similar to the one of J1714+6024 \citep{Marchesi_2019}. This small covering factor is also supported by the SED fitting results in Section~\ref{sec:SED_fitting}. We note that using a larger covering factor does not alter the best-fit values of other physical properties significantly.}

Both the phenomenological model and physically motivated models, i.e., {\tt MYTorus} and {\tt borus}, can well fit the source spectral shape. The best-fit results of the fluxes, line-of-sight column densities, and photon indices of all models are consistent with each other { within uncertainties. All models suggest that J1714+6027 is a Compton-thin quasar even after considering the uncertainties.} The residuals of both the phenomenological model and the physical model are consistent with each other, suggesting that the source does not present a significant reflection contribution from the torus. 

We note that future hard X-ray telescopes, e.g., HEX-P (https://hexp.org), could provide better characterization of the reflection component of the obscured quasars thanks to their much better sensitivities and photon collecting capabilities at hard X-rays \citep{Boorman2023}, thus better constraining the torus properties of quasars like J1714+6027. 

\subsection{Re-analysis of the 2011 XMM-Newton Observation} \label{sec:previous}
\begin{figure} 
\centering
\includegraphics[width=.48\textwidth]{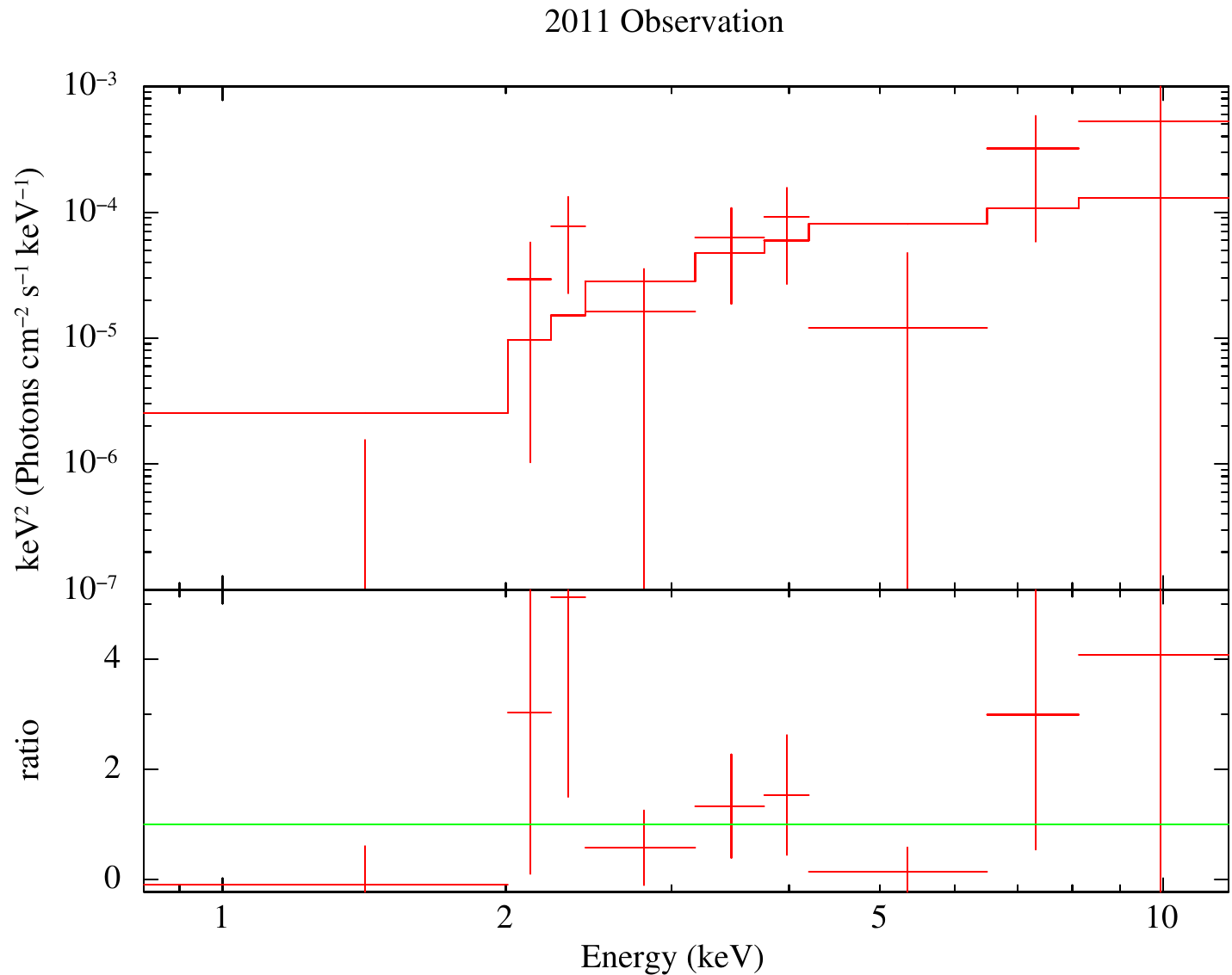}
\caption{The 2011 \XMM\ observed X-ray spectrum (in observed frame) of J1714+6027 fitted using phenomenological model described in Section~\ref{sec:phe}. The spectra are grouped with a minimum of 2 counts per bin. The red points correspond to the data and the solid line is the best-fit model prediction.}
\label{fig:2011_spectrum}
\end{figure}

\begin{figure} 
\centering
\includegraphics[width=.48\textwidth]{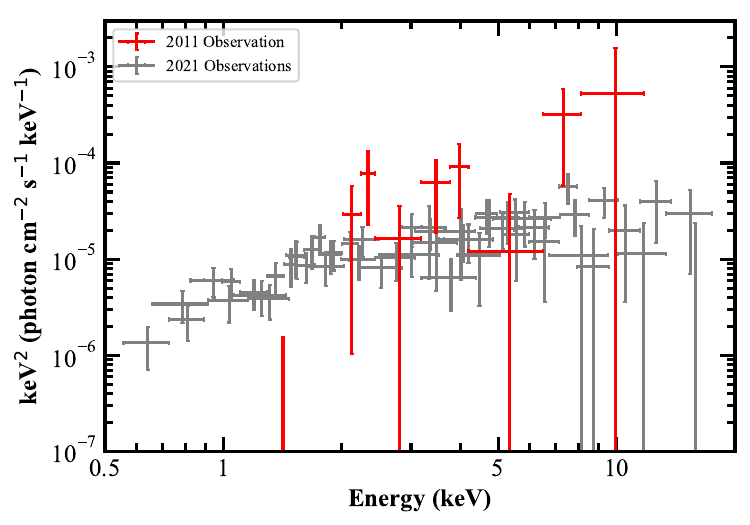}
\caption{Unfolded spectra of the \XMM\ observation in 2011 (red) and \XMM\ + \NuSTAR\ observations in 2021 (grey). Both the 2011 and 2021 spectra are modeled using the phenomenological model. The 2011 observation is modeled assuming a photon index of 1.45.}
\label{fig:spectrum_compare}
\end{figure}

\begin{figure} 
\centering
\includegraphics[width=.48\textwidth]{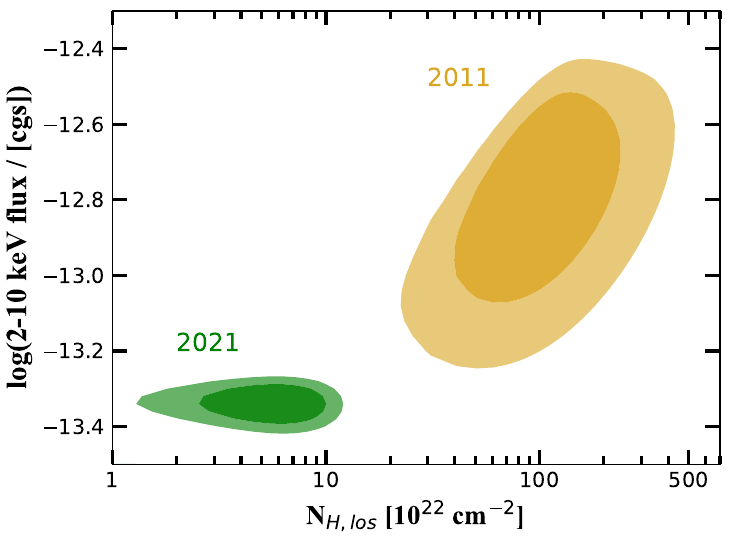}
\caption{Contour of the 2--10~keV flux as a function of the line-of-sight column density of the 2011 (yellow) and 2021 (green) observations. We present the fitting results from the phenomenological model of both observations. The dark and shallow colors represent the 68~\% and 90\% confidence levels, respectively.}
\label{fig:contour}
\end{figure}

As a consistency check, we re-fit the 2011 \XMM\ observation that led to a CT estimate of the source l.o.s. column density, this time using the photon index derived from the phenomenological model in the recent deeper observations ($\Gamma$ = 1.45). The spectrum and the best-fit model are plotted in Fig.~\ref{fig:2011_spectrum}. The best-fit `line-of-sight' column density is N$\rm _{H,Z}$ = 9.5$_{-5.9}^{+17.2}$ $\times$ $10^{23}$\,cm$^{-2}$. The best-fit flux of the source in 2011 observation is F$_{0.5-2}$ $<$ 1.2 $\times$ 10$^{-14}$ erg~cm$^{-2}$~s$^{-1}$ and F$_{2-10}$ = 1.7$_{-0.9}^{+1.5}$ $\times$ 10$^{-13}$ erg~cm$^{-2}$~s$^{-1}$. We compare the source spectra of the 2011 observation and the 2021 observations in Fig.~\ref{fig:spectrum_compare}. We plot the contour of the 2--10~keV flux as a function of the line-of-sight column density of the 2011 and 2021 observations in Fig.~\ref{fig:contour}. The best-fit intrinsic luminosity of the source in the 2011 observation is L$_{2-10}\sim$9 $\times$ $10^{45}$\,erg\,s$^{-1}$, which is about 5 times more luminous than what was measured in 2021. Therefore, the source might have experienced a significant variability in intrinsic luminosity as well. 

%The 1$\sigma$ uncertainties of the `line-of-sight' column density in 2011 and 2021 are N$\rm _{H,Z,2011}$ = 9.5$_{-3.9}^{+7.0}$ $\times$ $10^{23}$\,cm$^{-2}$ and N$\rm _{H,Z,2021}$ = 5.9$_{-2.3}^{+2.6}$ $\times$ $10^{22}$\,cm$^{-2}$, respectively. Therefore, the `line-of-sight' column density variability between 2011 and 2021 is at $\sim$2.3~$\sigma$. The 1$\sigma$ uncertainties of the 2--10~keV flux in 2011 and 2021 are F$_{2-10,2011}$ = 1.7$_{-0.6}^{+0.9}$ $\times$ 10$^{-13}$ erg~cm$^{-2}$~s$^{-1}$ and F$_{2-10,2021}$ = 4.6$_{-0.4}^{+0.4}$ $\times$ 10$^{-14}$ erg~cm$^{-2}$~s$^{-1}$, respectively. Therefore, the 2--10~keV flux variability between 2011 and 2021 is at $\sim$2.1~$\sigma$.

\section{Multiwavelength Analysis} \label{sec:multi}
J1714+6027 has rich multiwavelength data spanning from X-ray to radio. We report the multiwavelength data and SED fitting of the source in this section.

\begingroup
\renewcommand*{\arraystretch}{1.1}
\begin{table}[htb!]
\centering
\caption{Multiwavelength photometries of J1714+6027. We report 1\,$\sigma$ flux errors.}
\label{Table:SED}
  \begin{tabular}{lc}
       \hline
       \hline       
      	Band&Flux Density\\
       \hline
	2--10~keV$^1$ ($10^{-14}$\,erg\,cm$^{-2}$ s$^{-1}$)&3.36$\pm$0.44\\ 
	0.5--2 keV$^1$ ($10^{-14}$\,erg\,cm$^{-2}$ s$^{-1}$)&1.58$\pm$0.22\\ 
	6~keV$^{2}$ (10$^{-6}$ mJy)&1.74$\pm$0.23\\ 
	1.25~keV$^{2}$ (10$^{-6}$ mJy)&4.36$\pm$0.61\\
	DESI Image {\it g} (AB mag)&21.77$\pm$0.02\\	
	DESI Image {\it r} (AB mag)&21.20$\pm$0.02\\	
	DESI Image {\it z} (AB mag)&20.49$\pm$0.02\\	
	SDSS {\it u} (AB mag) &$>$23.38\\
	SDSS {\it g} (AB mag) &21.95$\pm$0.07\\
	SDSS {\it r} (AB mag) &21.49$\pm$0.07\\
	SDSS {\it i} (AB mag) &21.02$\pm$0.07\\	
	SDSS {\it z} (AB mag) &21.35$\pm$0.38\\	
	IRTF $\it H$ (AB mag)&19.69$\pm$0.09\\
	Spitzer  3.6~$\mu$m ($\mu$Jy)&178$\pm$19\\
	Spitzer  4.5~$\mu$m ($\mu$Jy)&248$\pm$26\\
	Spitzer  5.8~$\mu$m ($\mu$Jy)&495$\pm$55\\
	Spitzer 8.0~$\mu$m (mJy)&1.08$\pm$0.11\\
	WISE 3.4~$\mu$m ($\mu$Jy)&144$\pm$5\\
	WISE 4.6~$\mu$m ($\mu$Jy)&266$\pm$9\\
	WISE 12~$\mu$m (mJy)&2.15$\pm$0.08\\
	WISE 22~$\mu$m (mJy)&6.29$\pm$0.50\\
	Spitzer 24~$\mu$m (mJy)&5.60$\pm$0.07\\
	Spitzer 70~$\mu$m (mJy)&26.4$\pm$5.4\\
	Spitzer 160~$\mu$m (mJy)&59.0$\pm$20.4\\
	Herschel 250~$\mu$m (mJy)&$<$18\\
	Herschel 350~$\mu$m (mJy)&$<$15\\
	VLA 1.4 GHz (mJy)&1.23$\pm$0.06\\
	GMRT 610 MHz (mJy)&2.16$\pm$0.25\\
       \hline
	\hline
\end{tabular}
\tablecomments{$^1$The X-ray flux is corrected by the absorption requested by Cigale. %$^2$SDSS DR 18, taken in April 2020. PSF magnitude.
 $^{2}$Converted to Cigale format following \citet{Yang2020} eq.~1.}
\end{table}
\endgroup

\subsection{Multiwavelength Data}
J1714+6027 lies in the Spitzer Space Telescope \citep{Werner2004} Extragalactic First Look Survey \citep[XFLS,][]{Lacy2005,Fadda2006} area. The optical (rest-frame UV) spectrum of J1714+6027 was taken using the Palomar 200-inch telescope in 2005, which showed that J1714+6027 is a type 2 quasar at $z$ = 2.99 \citep{Lacy2007}. This is further confirmed by its near-infrared (rest-frame optical) spectrum taken using the Infrared Telescope Facility (IRTF) in 2007 \citep{Lacy2011}, which presented a clear O III emission line. J1714+6027 has wealthy multiwavelength data, including the photometry data from Dark Energy Spectroscopic Instrument (DESI) Legacy Imaging Surveys DR 9 \citep{Dey2019}, Sloan Digital Sky Survey (SDSS) DR 18, IRTF \citep{Lacy2011}, Wide-field Infrared Survey Explorer \citep[WISE,][]{Cutri2021}, Spitzer \citep{Lacy2005,Fadda2006,Frayer2006,Lacy2011}, Herschel \citep{Lacy2011}, Very Large Array \citep[VLA,][]{Condon2003}, Giant Metrewave Radio Telescope \citep[GMRT,][]{Garn2007}, and the X-ray data reported in this work. We list the multiwavelength data of J1714+6027 in Table~\ref{Table:SED}. 

\citet{Lacy2011} performed spectral energy distribution (SED) fitting of J1714+6027 using the photometry data from SDSS, IRTF, Spitzer, and Herschel. They used a host-galaxy stellar burst component (assuming dual stellar populations), an AGN torus component (assuming a power law with a cutoff at the dust sublimation temperature of 1500~K and a longer wavelength cutoff at $\sim$20~$\mu$m), and a cold dust component. \citet{Lacy2011} found different source star formation rates (SFRs) using different methods: specifically, they measured an SFR$\sim$130~$M_\sun$~yr$^{-1}$ using the UV luminosity, and SFR$\sim$4,000--5,000~$M_\sun$~yr$^{-1}$ using the far-infrared and radio luminosity, under the assumption that the emission in these bands is dominated by stellar emission. The authors argued that the anomalously huge SFR derived from the far-infrared and radio emission suggests that the flux in these two bands is heavily dominated by the AGN torus. This is also suggested by the lack of polycyclic aromatic hydrocarbon (PAH) features in the source mid-infrared 5--38~$\mu$m spectrum observed by Spitzer, as well as from the high dust temperature $\sim$135~K obtained from the SED fitting \citep{Lacy2011}. 
\begingroup
\renewcommand*{\arraystretch}{1.1}
\begin{table}[htb!]
\centering
\caption{{ Bayesian-like estimated values of the physical properties of J1714+6027 fitted with CIGALE.}}
\label{Table:CIGALE_result}
  \begin{tabular}{lccl}
       \hline
       \hline       
      	Property&CIGALE&\citet{Lacy2011}&Unit\\
       \hline
       	$\chi^{2}_{red}$&1.2&4.1\\
	SFR&1280 $\pm$ 77&130--5130&M$_\sun$~yr$^{-1}$\\
	$L_{\rm AGN}$$^1$&3.5 $\pm$ 0.2&2.5&10$^{13}$ $L_\sun$\\
	$L_{\rm stellar}$&1.6 $\pm$ 0.1&&10$^{13}$ $L_\sun$\\
%	$L_{\rm dust}$&1.0 $\pm$ 0.1&&10$^{13}$ $L_\sun$\\
	M$_{\rm star}$&1.1 $\pm$ 0.2&2.5&10$^{12}$ M$_\sun$\\
	M$_{\rm dust}$$^2$&$<$12&0.1&10$^{8}$ M$_\sun$\\
	M$_{\rm gas}$&6.5 $\pm$ 1.3&&10$^{11}$ M$_\sun$\\
	M$_{\rm gas,mo}$$^3$&&$<$2.8&10$^{10}$ M$_\sun$\\
	$\alpha_{\rm OX}$&--1.63 $\pm$ 0.05&\\
	RL$_{\rm 30deg}$$^4$&9.9 $\pm$ 0.6&\\
       \hline
	\hline
\end{tabular}
\tablecomments{$^1$ The sum of the observed AGN disk and dust reemitted luminosity. $^2$ The mass of the dust in the host galaxy. $^3$ Molecular gas mass derived converted from CO(1--0) luminosity of the source measured using Extended Very Large Array (EVLA). $^4$ Radio loudness at 30$^\circ$ inclination.}
\end{table}
\endgroup

\subsection{SED Fitting}\label{sec:SED_fitting}
Type 2 quasars constitute a less-biased sample to study the properties of the host galaxy of AGN as their rest-frame optical spectra are dominated by their host galaxies rather than the central nucleus. J1714+6027 is well-suited for such a study thanks to its wealthy multiwavelength data available. We performed our SED fitting using the Code Investigating GALaxy Emission (CIGALE) code \citep{Burgarella2005,Noll2009,Boquien2019}, which has been widely used to model the physical properties of galaxies with or without an active nucleus \citep[e.g.,][]{Wang2019,Florez2020,Salim2020,Masoura2021,Shirley2021,Thorne2021,Haro2023,Yang2023}. The X-ray data can be used to better constrain the AGN contribution to the rest-frame UV/optical spectrum using the well-established $\alpha_{\rm OX}$ relation \citep[e.g.,][]{Steffen2006,Just2007,Lusso2017}. Therefore, we used the most updated version of CIGALE v2022.1, which included an X-ray module to better constrain the contribution from AGN to the entire spectrum \citep{Yang2020,Yang2022}. The new version of CIGALE also updated the AGN model to account for a clumpy two-phase torus.

We followed the CIGALE configuration used in \citet{Yang2020,Yang2022} when analyzing J1714+6027. The model includes a stellar emission module, a Galactic dust emission module, and an AGN module. In the stellar emission component, we adopt a delayed star formation history (SFH), {\tt sfhdelayed}, and a single stellar population (SSP) assuming the \citet{Bruzual2003} model, {\tt bc03}. The nebular templates are based on \citet{Inoue2011}. We adopt the Galactic dust attenuation (GDA) module, {\tt dustatt\_calzleit} \citep{Calzetti2000,Leitherer2002}, and the dust emission {\tt dl2014} module \citep{Draine2014}.
The AGN emission is modeled by {\tt SKIRTOR} \citep{Stalevski2012,Stalevski2016}. The X-ray luminosity is connected to the accretion disk 2500\AA\ luminosity through $\alpha_{\rm OX}$.

\begin{figure} 
\centering
\includegraphics[width=.48\textwidth]{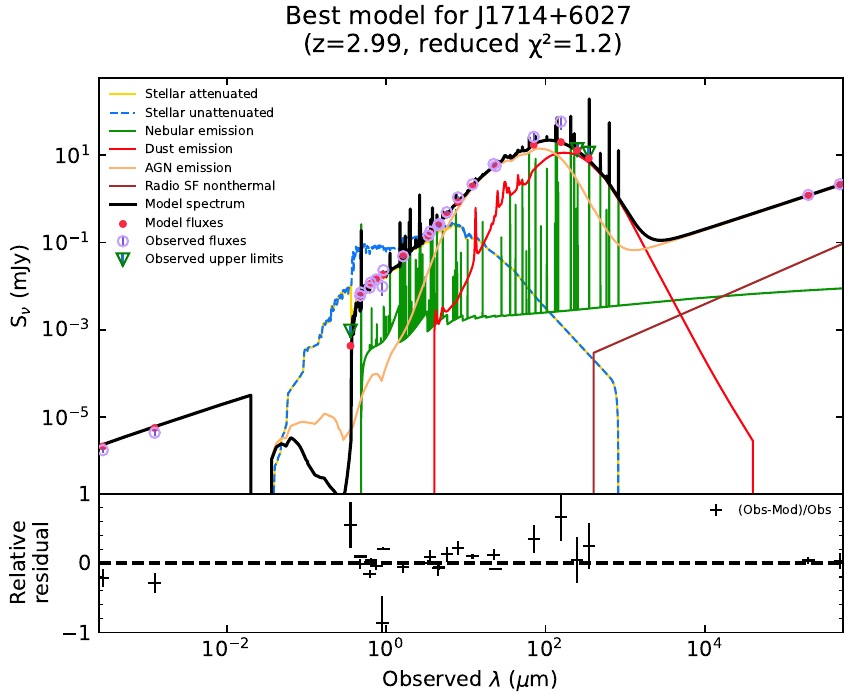}
\caption{Best-fit SED fitting result from CIGALE and relative residual. The line styles for each module are listed in the key. Relative residual is defined as ({\tt Obs}--{\tt Mod})/{\tt Obs}, where {\tt Obs} is the observation data and {\tt Mod} is the model prediction.}
\label{fig:SED}
\end{figure}

The SED with the best-fit models is plotted in Fig.~\ref{fig:SED}. We note that the AGN emission dominates the entire spectrum of J1714+6027, except for the rest-frame UV/optical band, which is dominated by stellar emission from the host galaxy. This is because of the obscuration along the line-of-sight of the AGN. The best-fit physical properties of the AGN and its host-galaxy of J1714+6027 are presented in Table~\ref{Table:CIGALE_result}. 

The detailed fitting strategy is discussed in \ref{sec:CIGALE_setup}, where we report the parameters used when fitting the SED in Table~\ref{Table:CIGALE}.

\section{Discussion} \label{sec:discuss}
We discuss the X-ray and multiwavelength variability of the source in Section~\ref{sec:X_var} and Section~\ref{sec:multi_var}. We then discuss the potential origin of the obscuration of the source observed in 2011 and 2021 in Section~\ref{sec:obscu}. We discuss the SED shape of the source in Section~\ref{sec:X_bol} and Section~\ref{sec:X_mid}. The source is an obscured hyperluminous IR galaxy, which is further discussed in Section~\ref{sec:DOG}. The source has a huge SFR and stellar mass, which is discussed in Section~\ref{sec:SFR}.
\subsection{X-ray Variability} \label{sec:X_var}
AGN are variable by definition \citep[e.g.,][]{Ulrich1997}. X-ray variability has been found in AGN in different time scales from seconds to years \citep[e.g.,][]{Fabian2009,Vagnetti2011,Yang2016,Middei2017,Paolillo2023}. However, simultaneous studies of X-ray flux and spectral shape variability have been performed on a limited number of high-$z$ quasars. J1714+6027 was found to be variable in its flux, column density, and intrinsic luminosity. In this section, we perform comprehensive simulations to further confirm the variability found in J1714+6027.
\begin{figure} 
\centering
\includegraphics[width=.48\textwidth]{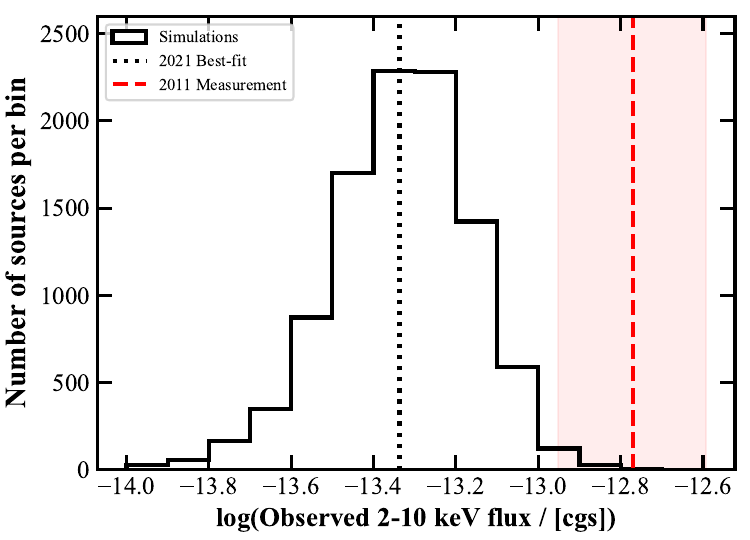}
\includegraphics[width=.48\textwidth]{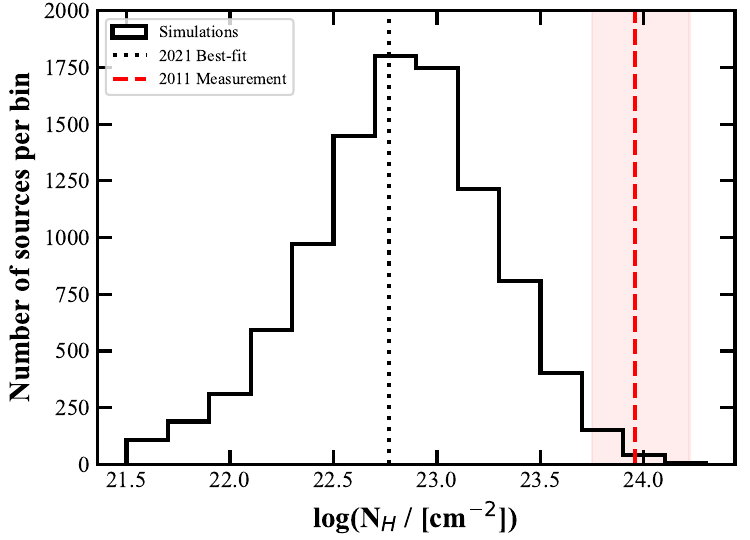}
\includegraphics[width=.48\textwidth]{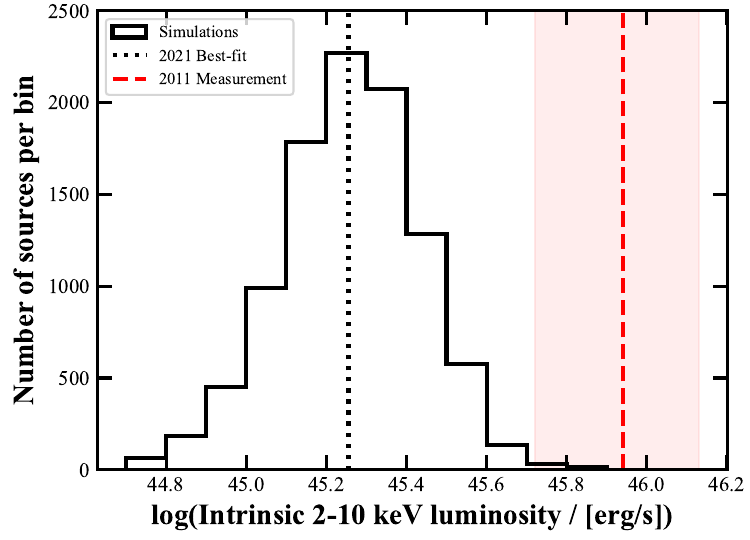}
\caption{Distributions of the 2--10~keV flux, column density, and 2--10~keV intrinsic luminosity measured in the 10,000 simulations. The black dotted line is the 2021 best-fit source properties. The red dashed line and the red shaded area are the 2011 measured source properties and its 1~$\sigma$ uncertainty, respectively.}
\label{fig:simulations}
\end{figure}   

Due to the limited number of source net counts of the 2011 observation, we performed a set of simulations to validate the X-ray variability found between the 2011 and 2021 observations. In detail, we simulated 10,000 spectra using the exposure and response files from the 2011 observation assuming the source properties were measured using the phenomenological model in 2021. In detail, we sample from the full co-variance associated with the fits of the 2011 observation using {\tt AllModels.simpars} function in PyXspec. We then fit each spectrum using the same model used to fit the 2011 observation with a photon index fixed at $\Gamma$ = 1.45. Therefore, if a non-negligible fraction of the best-fit results of the simulations were found to fall in the heavily obscured regime as measured in the 2011 observation, one cannot confirm that the source experienced significant variability, and instead assume that the heavily obscured measurement could have been due to the low photon statistics of the 2011 spectrum.

The median net count of the 10,000 simulated spectra is 10 counts, which is a little less than the 12 counts measured in the 2011 observation. The median 2--10~keV flux of the 10,000 simulated observations is F$_{2-10}$ = 4.8 $\times$ $10^{-14}$\,erg\,cm$^{-2}$ s$^{-1}$, which is consistent with the input F$\rm_{2-10}$ = 4.6 $\times$ $10^{-14}$\,erg\,cm$^{-2}$ s$^{-1}$ flux. We note that a fraction of 0.7\% of the simulated spectrum presents a flux falling in the range of 2--10~keV flux range measured in 2011 within 1~$\sigma$ confidence level, e.g., F$\rm_{2-10}$ = 0.8--3.2 $\times$ $10^{-13}$\,erg\,cm$^{-2}$ s$^{-1}$. The median column density of the 10,000 simulated spectra is N$\rm _{H,Z}$ = 6.9 $\times$ $10^{22}$\,cm$^{-2}$. We note a fraction of 1.4\% the simulated spectra having column densities falling in the range of the column density range measured in 2011 within 1~$\sigma$ confidence level, i.e., N$\rm _{H,Z}$ = 5.6--16.5 $\times$ $10^{23}$\,cm$^{-2}$. The median intrinsic 2--10~keV luminosity of the 10,000 simulated spectra is L$\rm_{2-10}$ = 1.9 $\times$ $10^{45}$\,erg\,s$^{-1}$, which is consistent with the input L$\rm_{2-10}$ = 1.8 $\times$ $10^{45}$\,erg\,s$^{-1}$ luminosity. We note a fraction of 0.3\% the simulated spectra having intrinsic 2--10~keV luminosity falling in the range of the intrinsic 2--10~keV luminosity range measured in 2011 within 1~$\sigma$ confidence level, i.e., N$\rm _{H,Z}$ = 5.2--13.4 $\times$ $10^{45}$\,erg\,s$^{-1}$.

We plot the distributions of the 2--10~keV flux, column density, and luminosity measured in the 10,000 simulations in Fig.~\ref{fig:simulations}. Therefore, the 2011 measured source flux, column density, and intrinsic luminosity are not consistent with those measured in the 2021 observations at $\sim2-3\,\sigma$ confidence level.

\subsection{Multiwavelength Variability} \label{sec:multi_var}
J1714+6027 presented significant X-ray flux variability between the 2011 observation and the 2021 observations. This flux variability is caused by both `line-of-sight' column density variability and intrinsic luminosity variability as shown in Section~\ref{sec:previous}. We then searched for potential multiwavelength variability of the source as hinted by the significant X-ray variability. We find a $\sim$0.2~mag difference in the (observed-frame) optical band between the SDSS 2000 observation ($m_r\sim$21.49~mag) and DESI 2015--2019 observations ($m_r\sim$21.20~mag). However, the optical flux of the source was consistent between 2009 and 2021 measured by the Palomar Transient Factory (PTF) and Zwicky Transient Facility (ZTF) with a higher average flux than both SDSS and DESI measurement (e.g., $m_r\sim$21.05~mag), although a larger flux uncertainty ($\sim$0.2~mag) of PTF and ZTF measurements is noted. Therefore, we argue that no significant flux variability was observed in the optical band between the two X-ray observations in 2011 and 2021. The WISE data also suggest a consistent near-IR flux of the source between 2010 and 2022 within uncertainties, although there was an observation gap between 2011 and 2013. Therefore, significant flux variability of J1714+6027 was only observed in the X-rays rather than in the optical and near-IR. This might be explained by the fact that the observed optical to near-IR (UV and optical in the rest-frame) spectra of J1714+6027 are dominated by the emission from the host galaxy (see Fig.~\ref{fig:SED}), which is much more stable compared to the nucleus in years scale.

\subsection{Origin of J1714+6027's Obscuration} \label{sec:obscu}
The line-of-sight column density of J1714+6027 in 2021 is a few times of 10$^{22}$~cm$^{-2}$. The origin of this obscuration can be either from the gas in the broad line region (BLR, $<$0.1~pc), from the dusty gas in the torus ($<$10~pc), or the interstellar medium ($>$kpc) in its host galaxy \citep[see][for a review]{Hickox18}. Recent works \citep[e.g.,][]{Gilli2022} showed that the column density of ISM in the host galaxy increases rapidly toward higher redshift using deep ALMA observations. Their works showed that ISM could provide an obscuration up to 3 $\times$ 10$^{23}$~cm$^{-2}$ column density at $z\sim3$, suggesting that the obscuration of J1714+6027 observed in 2021 can be caused by the ISM in its host galaxy considering its high SFR, although we cannot exclude the possibility that the obscuration is caused by the torus.

However, the ISM in the host galaxy alone cannot explain the obscuration found in the 2011 observation of J1714+6027. This is because not only the column density observed in 2011 is larger than 3 $\times$ 10$^{23}$~cm$^{-2}$, but also galactic obscuration should not vary in $\le$2.5 (source rest-frame) years. Therefore, the AGN circumnuclear material has to be involved in the obscuration found in 2011. Considering the significant variability found in equal to or less than 10 years (or 2.5 years in the source rest-frame), the obscuration could be from either the BLR or the torus. 

Here we calculate the size of the obscurer by multiplying the velocity of the obscurer ($V_{\rm obs}$) and the variability time (2.5 rest-frame years). Assuming the motion of the obscurer is dominated by the gravitational field of the central SMBH, the velocity of the obscurer is $V_{\rm obs}$ = ($G$M$_{\rm BH}$/$r$)$^{1/2}$, where $G$ is the gravitational constant, M$_{\rm BH}$ is the mass of the SMBH, and $r$ is the distance between the obscurer and the central SMBH. If the obscurer is in the BLR, the size of the obscurer is $>$ 0.022~pc (assuming that BLR is 0.1~pc away from the SMBH and the variability happened in source rest-frame 2.5 years). If the obscurer is in the torus (assuming at 1~pc away from the SMBH), the obscurer size is $>$ 0.007~pc. Here we assume that the SMBH of J1714+6027 is accreting materials less than the Eddington limit. The bolometric luminosity of J1714+6027 is 2.16 $\times$ 10$^{47}$~erg~s$^{-1}$ (see Section~\ref{sec:X_bol}), so the black hole mass is M$_{\rm BH}\ge$ 1.7 $\times$ 10$^{9}$ M$_\sun$. Monitoring the line-of-sight column density variability of the source could provide more information on the properties and distribution of the obscurer surrounding the SMBH. Nevertheless, the fast line-of-sight column density variability suggests a very dynamic environment surrounding the SMBH and the circumnuclear materials could be clumpy.

\begin{figure} 
\centering
\includegraphics[width=.48\textwidth]{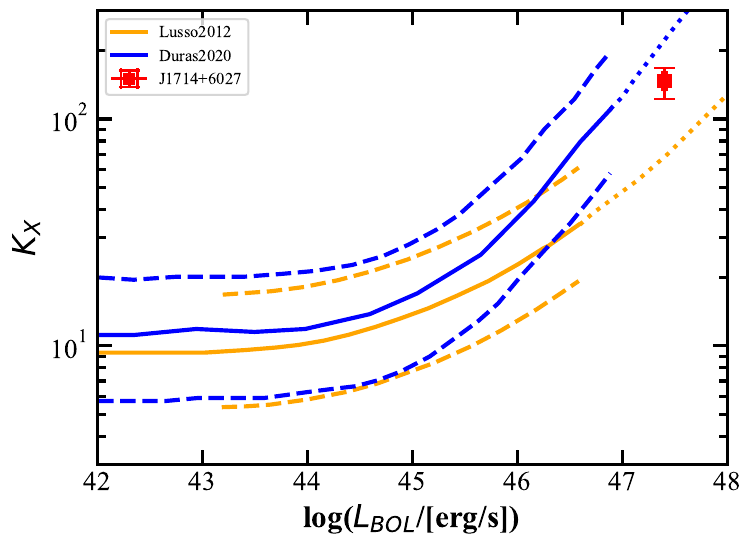}
\caption{$K_{\rm X}$ as a function of bolometric luminosity of type 2 AGN. The red square plot is the J1714+6027 (the error bar is in 1\,$\sigma$). The orange and blue lines are the measurements (solid) and extrapolation (dotted) derived in \citet{Lusso2012} and \citet{Duras2020}, respectively. The dashed lines correspond to the 1$\sigma$ spread of the samples used in the two works. The bolometric luminosities of the type 2 AGN in their work span a range from 10$^{42}$~erg~s$^{-1}$ to 2 $\times$ 10$^{46}$~erg~s$^{-1}$. Beyond these bolometric luminosities, the $K_{\rm X}$-correction was extrapolated from the measured data following $K_{\rm X}$ = a[1+(log($L_{\rm BOL}$/$L_\sun$)/b)$^c$]. $a,~b,~c$ are coefficients that were measured to be constants as a function of the bolometric luminosity.}
\label{fig:k_correction}
\end{figure}

\subsection{X-ray to Bolometric Luminosity of AGN} \label{sec:X_bol}

%The AGN disk and dust reemitted luminosity is 1.15 $\times$ 10$^{47}$~erg~s$^{-1}$ (Table~\ref{Table:CIGALE_result}). Therefore, the AGN bolometric luminosity of J1714+6027 is 1.25 $\times$ 10$^{47}$~erg~s$^{-1}$. We note that the radio luminosity contributes only 0.04\% to the entire AGN luminosity. The X-ray luminosity is 8.7\% of the AGN bolometric luminosity and the 2--10~keV X-ray luminosity is 1.3\% of the AGN bolometric luminosity. We note that the radio luminosity contributes only 0.14\% to the entire AGN luminosity. The X-ray luminosity is 29\% of the AGN bolometric luminosity and the 2--10~keV X-ray luminosity is 4.4\% of the AGN bolometric luminosity. 

The $K_{\rm X}$-correction, i.e., the 2--10~keV luminosity to the bolometric luminosity ($L_{\rm BOL}$) relation, is a significant ingredient to study the energy transfer process in AGN, which was found to significantly depend on the bolometric luminosity but not depend on the redshift \citep[e.g.,][]{Duras2020}. The $K_{\rm X}$-correction has been widely used to derive the bolometric luminosity of the X-ray detected AGN. \citet{Lusso2012} and \citet{Duras2020} derived $K_{\rm X}$ as a function of bolometric luminosity using a sample of X-ray-detected AGN. They found a very similar $K_{\rm X}$-$L_{\rm BOL}$ relation for type 1 AGN, but the $K_{\rm X}$-$L_{\rm BOL}$ relation in the luminous end of the type 2 AGN was not well constrained (Fig.~\ref{fig:k_correction}) due to the lack of luminous type 2 AGN sample.

J1714+6027 is a good target for constraining the $K_{\rm X}$-$L_{\rm BOL}$ relation at the luminous end. We used the 2--10~keV intrinsic luminosity of source in the 2021 observation derived from the {\tt borus} model, which is 1.7 $\times$ 10$^{45}$~erg~s$^{-1}$. As the bolometric luminosity of type 2 AGN cannot be derived directly from integrating the AGN emission components in the SED fitting (from X-ray to optical bands), we computed the bolometric luminosity of J1714+6027 following the method used in \citet{Lusso2012,Duras2020}. In their work, the bolometric luminosity of type 2 AGN was derived by scaling the IR (from 1~$\mu$m to 1000~$\mu$m) luminosity of the AGN by a factor of 1.9 \citep{Pozzi2007}. Therefore, the bolometric luminosity of J1714+6027 is $L_{\rm BOL}$ = 2.5 $\times$ 10$^{47}$~erg~s$^{-1}$. Therefore, the $K_{\rm X}$-correction of J1714+6027 is $K_{\rm X}$ = 147.

Fig.~\ref{fig:k_correction} plots $K_{\rm X}$ as a function of the bolometric luminosity of type 2 AGN derived in \citet{Lusso2012} and \citet{Duras2020}. We note that J1714+6027 has the largest bolometric luminosity among the sources in both of the two samples. J1714+6027 lies between the $K_{\rm X}$--$L_{\rm BOL}$ relations extrapolated from the two samples, but is closer to the more recent \citet{Duras2020} derived relation. It is shown that J1714+6027 is a valuable source to derive the intrinsic $K_{\rm X}$--$L_{\rm BOL}$ relation, but a larger sample of such luminous type 2 quasar is needed to constrain the relation.

The above correlation suggested that the X-ray emission contributes much less at the bright end. We note that the optical to bolometric luminosity ratio ($K_{\rm O}$) is nearly a constant as a function of $L_{\rm BOL}$, even at the bright end \citep{Duras2020}. Therefore, this might imply that the X-ray emitter, i.e., the corona, might have different physical or geometrical properties in different luminosity regimes \citep[e.g.,][]{Zappacosta2020}.

\subsection{X-ray to Mid-IR Luminosity} \label{sec:X_mid}
\begin{figure} 
\centering
\includegraphics[width=.48\textwidth]{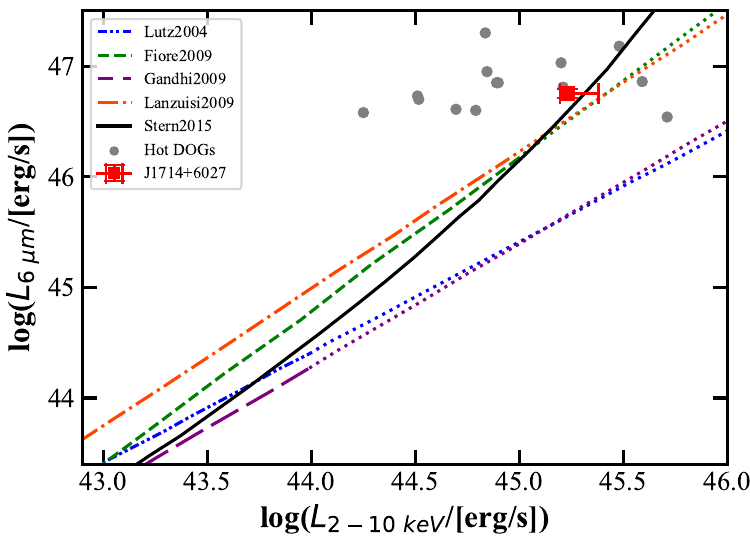}
\caption{Rest-frame 6~$\mu$m AGN luminosity as a function of the 2-10~keV intrinsic luminosity of J1714+6027 (red square) compared with the mid-IR to X-ray relation obtained by \citet[][blue dotted line]{Lutz2004}, \citet[][green short dashed line]{Fiore2009}, \citet[][purple long dashed line]{Gandhi09}, \citet[][orange dash-dot line]{Lanzuisi2009}, and \citet[][black solid line]{Stern2015}. The extrapolations of each work to higher luminosities are plotted in dotted lines. Hot DOGs from \citet{Assef2016,Ricci_2017,Vito2018,Zappacosta2018} are plotted as gray circles.}
\label{fig:6micron}
\end{figure}

The X-ray and mid-IR emission from AGN are strongly correlated due to the same energy budget. In the last decades, the intrinsic X-ray and mid-IR luminosities of AGN were explored in different redshift and luminosity scales \citep[e.g.,][]{Lutz2004,Fiore2009,Gandhi09,Lanzuisi2009,Asmus15,Stern2015}. Indeed, a strong correlation between the X-ray and mid-IR luminosity was detected. However, such a correlation was found to evolve with AGN luminosities: the X-ray luminosity increases slower than the mid-IR luminosity as the AGN luminosity increases \citep{Stern2015} and Fig.~\ref{fig:6micron}. This might suggest that there is an evolution of the AGN mid-IR emitter, in which the band is typically dominated by the AGN dusty torus. It is worth mentioning that the correlations obtained by \citet{Lutz2004} and \citet{Gandhi09} were both derived from local AGN samples, which could well separate the AGN and host-galaxy emissions but only sampled the low-luminosity regime ($L_{\rm 2-10~keV}\le10^{44}$~erg~s$^{-1}$). \citet{Fiore2009} and \citet{Lanzuisi2009} sampled more luminous quasars with $L_{\rm 2-10~keV}$ upto $10^{45}$~erg~s$^{-1}$. The brightest quasars with $L_{\rm 2-10~keV}$ upto $10^{46}$~erg~s$^{-1}$ were sampled by \citet{Stern2015}. But they are not able to resolve the nuclei.

We calculated the rest-frame mid-IR 6~$\mu$m and 2--10~keV intrinsic luminosity from the best-fit SED of J1714+6027. The object is well consistent with the correlations derived in \citet{Fiore2009}, \citet{Lanzuisi2009}, and \citet{Stern2015}. It is worth mentioning that the AGN mid-IR luminosity of the sources in their sample can have non-negligible contamination from the host galaxy even for type 1 AGN \citep{Yang2020}. Therefore, \citet{Fiore2009}, \citet{Lanzuisi2009}, and \citet{Stern2015} might overestimate the mid-IR luminosity of the AGN by including the contribution from the host galaxies. The entire 6~$\mu$m luminosity of J1714+6027 is $L_{\rm 6\mu m,tot}$ = 6.2 $\times$ $10^{46}$~erg~s$^{-1}$, which is 11\% higher than its AGN luminosity ($L_{\rm 6\mu m, AGN}$ = 5.6 $\times$ $10^{46}$~erg~s$^{-1}$) as plotted in Fig.~\ref{fig:6micron}. 

\begin{figure} 
\centering
\includegraphics[width=.48\textwidth]{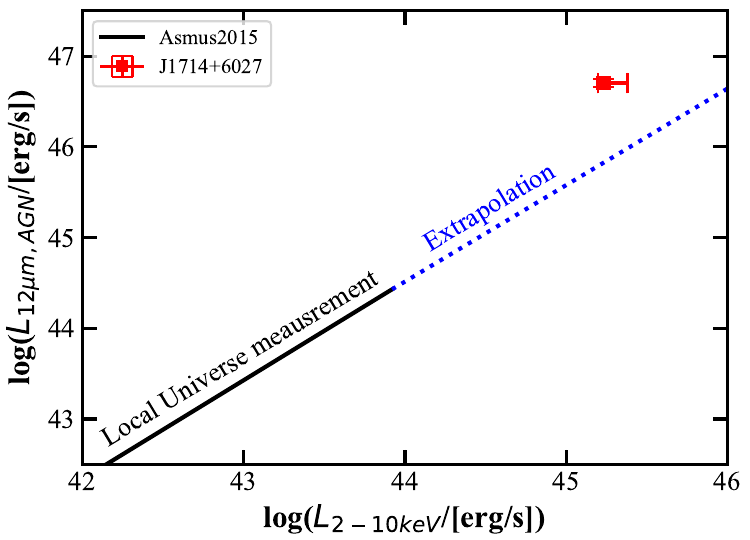}
\caption{Rest-frame 12~$\mu$m AGN luminosity as a function of 2-10~keV intrinsic luminosity of J1714+6027 (red square) and \citet{Asmus15} measurement using a sample of obscured (22$<$log(N$_H$)$<$23) AGN in the local Universe and its extrapolation to high luminosities (blue dots).}
\label{fig:12micron}
\end{figure}   

\citet{Asmus15} explored the 12~$\mu$m to X-ray luminosity correlation using a sample of mid-IR resolved AGN in the local Universe ($L_{\rm 2-10~keV}\le10^{44}$~erg~s$^{-1}$). We plot the correlation derived in \citet{Asmus15} with similar column densities (22$<$log N$_{\rm H}<$23) with what was measured in J1714+6027 in Fig.~\ref{fig:12micron}. The 12 $\mu$m luminosity of J1714+6027 is $L_{\rm 12\mu m, AGN}$ = 5.1 $\times$ $10^{46}$~erg~s$^{-1}$. We find a significant offset between the \citet{Asmus15} best-fit correlation and J1714+6027, suggesting an evolution of the AGN mid-IR emitter as found in the 6~$\mu$m to X-ray luminosity correlation.

The evolution of the X-ray to mid-IR correlation presented in Fig.~\ref{fig:6micron} and Fig.~\ref{fig:12micron} might be due to the evolution of the $K_{\rm X}$--$L_{\rm BOL}$ correlation as the mid-infrared photons are the UV/optical photons re-processed by the circumnuclear dust.

%\url{https://ui.adsabs.harvard.edu/abs/2017ApJ...835..105R/abstract}

\subsection{J1714+6027 and Hot Dusty-Obscured Galaxy} \label{sec:DOG}

The total IR (8--1000~$\mu$m) luminosity of J1714+6027 is $L_{IR}\sim$2.6 $\times$ 10$^{13}$~$L_\odot$, suggesting that it is an IR hyper-luminous galaxy \citep[$L_{IR}$ = 10$^{13}$--10$^{14}$~$L_\odot$, e.g.,][]{Eisenhardt2012}. The peak of the SED of J1714+6027 is at (source rest-frame) $\lambda\sim$21~$\mu$m, suggesting that the dust is hot \citep[$T\gg100~K$, derived with the relation presented in Fig.~20 in][]{Casey2014}. Its SED also presents a significant bump at the IR band, suggesting a large amount of dust inside the torus of the AGN. Therefore, J1714+6027 is an IR hyper-luminous, obscured galaxy with a significant amount of dust in the torus. 

In the last two decades, a sample of IR hyper-luminous, hot, dust-obscured galaxies (DOGs) were discovered \citep{Eisenhardt2012,Wu2012,Tsai2015}, selected using the W1W2 dropout criteria (i.e., these sources are bright in the WISE W3 and W4 band but are very faint (or not detected) in the W1 and W2 band). J1714+6027 does not fulfill the definition of hot DOGs, but they share similar IR luminosity and dust temperature. We note that hot DOGs might have stronger IR emission compared with their optical emission, which suggests that hot DOGs have a larger amount of dust in the torus or the central engine is more obscured by the dust. It is worth mentioning that dust-obscured galaxies (DOGs) were selected using the criteria with 24~$\mu$m flux density $F_{\rm 24\mu m}\ge$0.3~mJy and $F_{\nu}$(24~$\mu$m)/$F_{\nu}$(R)$\ge$1000 \citep{Dey2008}. J1714+6027 has a 24~$\mu$m flux density $F_{\rm 24\mu m}$ = 5.6~mJy but a little less $F_{\nu}$(24~$\mu$m)/$F_{\nu}$(R)$\sim$500, suggesting a lower IR to optical flux ratio. Nevertheless, it is worth comparing the X-ray properties of J1714+6027 with those of the hot, DOGs. 

The intrinsic X-ray luminosities of hot DOGs were found widely spread at similar IR luminosities \citep[see Fig.~\ref{fig:6micron},][]{Assef2016,Ricci_2017,Vito2018,Zappacosta2018}, with a large fraction of them are intrinsically X-ray weak compared with other AGN (especially type 1 AGN) at similar luminosities. However, it is worth noting that the obscurations of half of the hot DOGs in the \citet{Ricci_2017} sample (where most of the intrinsically X-ray weak hot DOGs were discovered) are derived from the empirical equation between $E(B-V)$ and column density (see Equation 1 in their paper). This method can lead to column density with large uncertainty, thus an uncertain intrinsic X-ray luminosity, especially in the heavily obscured regime. J1714+6027 presents a similar X-ray to mid-IR correlation as of other unobscured AGN (Section~\ref{sec:X_mid}) and is indistinguishable from the hot DOGs sample regarding the X-ray to mid-IR relation.

Hot DOGs are thought to be powered by highly obscured (N$_{\rm H}\gg$10$^{23}$~cm$^{-2}$), with many being CT, AGN \citep[e.g.,][]{Assef2016}, although the X-ray obscuration measurements of the current hot DOGs sample are highly uncertain \citep{Piconcelli2015,Ricci_2017,Vito2018,Zappacosta2018}. J1714+6027 is indeed less obscured in 2021 compared with hot DOGs. However, it is worth noting that J1714+6027 is an obscuration variable and was much more obscured in 2011. Therefore, J1714+6027 is indistinguishable from the hot DOGs sample regarding the X-ray obscuration.

 Therefore, we found that although J1714+6027 does not follow the definition of hot DOGs, it shares many similar properties with hot DOGs. There might be an evolutionary sequence of dust obscured, IR luminous sources as proposed in \citet{Assef2022}. 
%The AGN IR luminosity of J1714+6027 is 1.5 $\times$ 10$^{13}$~$L_\odot$.

\begin{figure}
\centering
\includegraphics[width=.48\textwidth]{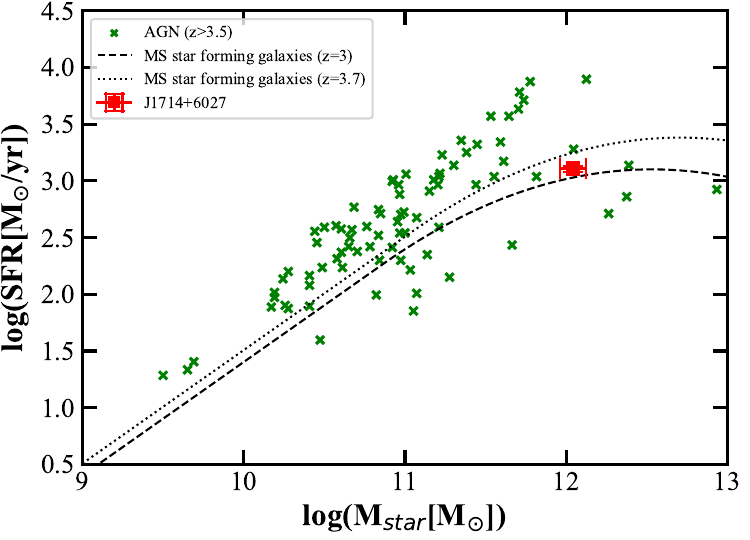}
\caption{SFR as a function of stellar mass of J1714+6027 (red square) and $z\ge$3.5 AGN (median redshift of 3.7) measured in \citet{Pouliasis2022}. The dashed and dotted lines represent the SFR and stellar mass correlations of $z$ = 3 and $z$ = 3.7 MS star-forming galaxies derived in \citet{Schreiber2015}, respectively.}
\label{fig:SFR}
\end{figure}   

\subsection{SFR of the Host-Galaxy of AGN in the early Universe} \label{sec:SFR}
The AGN feedback is thought to be involved in shaping the star forming process of their hots galaxies \citep[e.g.,][]{Granato2004,Di_Matteo2005,Fabian2012,Madau2014}. Type 2 quasars are great samples to extract the physical properties of the stellar population in the host galaxy of AGN as the stellar emission from the host galaxy is overwhelmed by the nuclei emission in type 1 AGN. Huge SFR ($\sim$1280~M$_{\sun}$~yr$^{-1}$) and stellar mass (M$_{\rm star}\sim$1.1 $\times$ 10$^{12}$~M$_{\sun}$) were found in the host-galaxy of J1714+6027, which were rare to be found in AGN even at this redshift \citep[e.g.,][]{Lanzuisi2017}. \citet{Pouliasis2022} studied the host-galaxy properties of a sample of X-ray detected, high-redshift ($z\ge$3.5) AGN. J1714+6027 is in the stellar active and massive end of the \citet{Pouliasis2022} sample, whose median SFR is $\approx$240~M$_{\sun}$~yr$^{-1}$ and median stellar mass is M$_{\rm star}\sim$5.6 $\times$ 10$^{10}$~M$_{\sun}$.The SFR as a function of the stellar mass of J1714+6027 is consistent with the \citet{Pouliasis2022} sample as shown in Fig.~\ref{fig:SFR}. 

The main-sequence (MS) star-forming galaxies also present a strong correlation between SFR and stellar mass \citep[e.g.,][]{Schreiber2015}. \citet{Pouliasis2022} found that a large fraction of the high-redshift AGN in their sample presented enhanced SFR compared with the high-redshift (median $z\sim$3.7) MS star-forming galaxies with similar stellar mass \citep{Schreiber2015}. However, the SFR is not enhanced by AGN compared with MS star-forming galaxies with M$_{\rm star}>$10$^{12}$~M$_{\sun}$, which is also evident by J1714+6027 (Fig.~\ref{fig:SFR}). This might suggest that the gas reservoir in the host galaxy rather than the presence of AGN regulates the SFR in the massive end of galaxies.

\section{Conclusions}

J1714+6027 is a type 2 quasar at $z$ = 2.99. Its 2011 shallow X-ray observation suggested that the source is a candidate CT-quasar, which was then reviewed with deep \NuSTAR\ and \XMM\ observations in the year 2021. In this work, we analyzed its recent X-ray data and we performed an SED fitting of the source thanks to its great multiwavelength data. We find that:

\begin{enumerate}
\item The source was found to be a candidate CT-quasar with a column density of N$\rm _{H,Z,2011}$ = 9.5$_{-5.9}^{+17.2}$ $\times$ $10^{23}$\,cm$^{-2}$. The recent deep \NuSTAR\ and \XMM\ observations showed that the source is less obscured with a column density of N$\rm _{H,Z,2021}$ = 5.9$_{-3.6}^{+4.9}$ $\times$ $10^{22}$\,cm$^{-2}$. 

\item The source presented significant variability in its X-ray flux, obscuration, and intrinsic luminosities at 2--3~$\sigma$ confidence level in 2.5 years (in source rest-frame). 

\item The source was not found to vary as significantly in the optical and near-IR bands as we observed in the X-rays. This might be because the source emission in the optical and near-IR bands is dominated by the host galaxy.

\item We explored the origin of the X-ray obscuration of the source. The previously measured heavy obscuration in the year 2011 can be from either the AGN BLR or the AGN torus. We estimated that the size of the obscurer is $>$0.022~pc if it is in the BLR or is $>$0.007~pc if it is in the torus. The recently measured lower obscuration of the source in the year 2021 can originate from the AGN torus and its host galaxy.

\item The source is an extremely luminous type 2 quasar with a bolometric luminosity of $L_{\rm BOL}$ = 2.5 $\times$ 10$^{47}$~erg~s$^{-1}$. The $K_{\rm X}$--$L_{\rm BOL}$ correlation is less constrained in the bright end of the bolometric luminosity. J1714+6027 has the largest bolometric luminosity among the sources in both \citet{Lusso2012} and \citet{Duras2020} samples. We found that J1714+6027 lies between the $K_{\rm X}$--$L_{\rm BOL}$ relations extrapolated from the two samples, but is closer to the more recent \citet{Duras2020} $K_{\rm X}$--$L_{\rm BOL}$ relation. A larger sample of bright AGN is needed to constrain the correlation at the bright end, which will also be vital to understanding the AGN corona at different luminosity.

\item We found that the source followed the evolution of the correlation between X-ray luminosity and mid-IR luminosity where a much higher mid-IR luminosity is found for high X-ray luminosity sources. This might be due to the evolution of the $K_{\rm X}$--$L_{\rm BOL}$ correlation.

\item The source is an IR hyper-luminous, obscured galaxy with a significant amount of hot dust in its torus. We found that J1714+6027 shares many similar properties with hot DOGs.

\item The source has a huge SFR ($\sim$1280~M$_{\sun}$~yr$^{-1}$) and stellar mass (M$_{\rm star}\sim$1.1 $\times$ 10$^{12}$~M$_{\sun}$). Its SFR and stellar mass correlation are consistent with what was found in other $z>$3.5 AGN. The correlation is also consistent with what was found in the $z$ = 3 main sequence star-forming galaxies, suggesting that the existence of an active nucleus does not enhance the star-formation rate.

\end{enumerate}

%\subsection{Importance of Far-IR Data}
%%%%%%%%%%%%%%%%%%

%%%%%%%%%%%%%%%%%%
\section{Acknowledgments}
XZ thanks the anonymous referee for their very helpful comments on the manuscript. XZ acknowledges NASA funding under contract number 80NSSC20K0043. XZ thanks Yue Shen and Roberto Assef for their very helpful comments on the paper. XZ thanks Qian Yang, Guang Yang, and Mingyang Zhuang for the helpful discussion on the SED fitting. 
We thank Karl Foster and Murry Brightman and the \NuSTAR\ observation planning team for their help in designing the observation plan and scheduling the observations. 

This work has made use of data from the \NuSTAR\ mission, a project led by the California Institute of Technology, managed by the Jet Propulsion Laboratory, and funded by the National Aeronautics and Space Administration. 
This research has made use of the \NuSTAR\ Data Analysis Software (NuSTARDAS) jointly developed by the ASI Science Data Center (ASDC, Italy) and the California Institute of Technology (USA). 
This research has made use of data and software provided by the High Energy Astrophysics Science Archive Research Center (HEASARC), which is a service of the Astrophysics Science Division at NASA/GSFC and the High Energy Astrophysics Division of the Smithsonian Astrophysical Observatory. 
This work is based on observations obtained with \XMM, an ESA science mission with instruments and contributions directly funded by ESA Member States and NASA. 
This work makes use of the data from SDSS. Funding for the Sloan Digital Sky Survey has been provided by the Alfred P. Sloan Foundation, the U.S. Department of Energy Office of Science, and the Participating Institutions. 
This work is partly based on the data from WISE, which is a joint project of the University of California, Los Angeles, and the Jet Propulsion Laboratory/California Institute of Technology.
The DESI Legacy Imaging Surveys consist of three individual and complementary projects: the Dark Energy Camera Legacy Survey (DECaLS), the Beijing-Arizona Sky Survey (BASS), and the Mayall z-band Legacy Survey (MzLS). DECaLS, BASS and MzLS together include data obtained, respectively, at the Blanco telescope, Cerro Tololo Inter-American Observatory, NSF’s NOIRLab; the Bok telescope, Steward Observatory, University of Arizona; and the Mayall telescope, Kitt Peak National Observatory, NOIRLab. NOIRLab is operated by the Association of Universities for Research in Astronomy (AURA) under a cooperative agreement with the National Science Foundation. Pipeline processing and analyses of the data were supported by NOIRLab and the Lawrence Berkeley National Laboratory (LBNL). Legacy Surveys was supported by: the Director, Office of Science, Office of High Energy Physics of the U.S. Department of Energy; the National Energy Research Scientific Computing Center, a DOE Office of Science User Facility; the U.S. National Science Foundation, Division of Astronomical Sciences; the National Astronomical Observatories of China, the Chinese Academy of Sciences and the Chinese National Natural Science Foundation. LBNL is managed by the Regents of the University of California under contract to the U.S. Department of Energy. 
This work is based in part on archival data obtained with the Spitzer Space Telescope, which was operated by the Jet Propulsion Laboratory, California Institute of Technology under a contract with NASA. Support for this work was provided by NASA.
Herschel is an ESA space observatory with science instruments provided by European-led Principal Investigator consortia and with important participation from NASA.
The National Radio Astronomy Observatory is a facility of the National Science Foundation operated under cooperative agreement by Associated Universities, Inc.
We thank the staff of the GMRT that made these observations possible. GMRT is run by the National Centre for Radio Astrophysics of the Tata Institute of Fundamental Research.

%%%%%%%%%%%%%%%%%%
%\facilities{\NuSTAR, \XMM}

%%%%%%%%%%%%%%%%%%
\appendix
\renewcommand{\thesection}{APPENDIX~\Alph{section}}
\section{X-ray spectral analysis of the sources near J1714+6027}\label{sec:near_source}
We analyze the \XMM\ spectra of three sources near J1714+6027, which are J171430+602722, J171430+602635, and J171425+602928. We extracted the \XMM\ spectra of the two sources following Section~\ref{sec:XMM}. The two sources are point sources in their optical images, so we fit their spectra with the phenomenological model as described in Section~\ref{sec:phe} assuming they are AGN. They do not have redshift information, so we assume they are at $z$ = 0. We found that their spectra can be well fitted with a simple power law model, without the need for an additional absorption component besides the galactic absorption. Therefore, we fit the spectra only with a simple power law model in case we overfit the data. The best-fit results are listed in Table~\ref{Table:best-fit_close}. 

 J171425+602928 is bright in \NuSTAR, so we analyze its \NuSTAR\ (3--24~keV) spectrum as well. We found that its best-fit photon index is $\Gamma$ = 1.75$^{+0.35}_{-0.37}$ and 2--10~keV flux is F$_{2-10}$ = 1.3$_{-0.3}^{+0.2}$ $\times$ $10^{-13}$\,erg\,cm$^{-2}$ s$^{-1}$, which is different from what we measured from its \XMM\ spectra. We note that this might be due to the variability of the source. We found that J171425+602928 presented evidence for intra-observation net count rate variability during the \XMM\ observations at $>5\sigma$ level. The net count rate varied by up to 5 times during the \XMM\ observations. As \XMM\ observation only overlapped with the \NuSTAR\ observation by 30\%, the source might have higher flux in the rest of the \NuSTAR\ observation when the source was not observed by \XMM. Nevertheless, detailed variability and spectral analysis are out of the scope of this paper. 

\begingroup
\renewcommand*{\arraystretch}{1.2}
\begin{table}
\centering

\caption{Summary of best-fit of \XMM\ spectra of J171430+602722, J171430+602635, and J171425+602928 using the phenomenological model.}
\label{Table:best-fit_close}
\scriptsize
  \begin{tabular}{lccc}
       \hline
       \hline       
       &J171430+602722&J171430+602635&J171425+602928\\
       \hline
       $C$/d.o.f.&20/33&20/29&183/171\\
       $\Gamma$&1.73$^{+0.20}_{-0.21}$&1.90$^{+0.23}_{-0.25}$&2.22$^{+0.06}_{-0.07}$\\
%       N$\rm _{H,Z}$\footnote{``line-of-sight'' column density in $10^{22}$\,cm$^{-2}$.}&5.9$_{-3.6}^{+4.5}$\\
       F$_{0.5-2}$\footnote{0.5--2\,keV flux in $10^{-14}$\,erg\,cm$^{-2}$ s$^{-1}$.}&1.1$_{-0.1}^{+0.2}$&1.0$_{-0.1}^{+0.1}$&8.2$_{-0.4}^{+0.3}$\\
       F$_{2-10}$\footnote{2--10\,keV flux in $10^{-14}$\,erg\,cm$^{-2}$ s$^{-1}$.}&2.1$_{-0.6}^{+0.6}$&1.4$_{-0.5}^{+0.5}$&7.3$_{-0.7}^{+0.7}$\\
       \hline
	\hline
\end{tabular}
\end{table}
\endgroup

\section{CIGALE Setup} \label{sec:CIGALE_setup}
We set up the CIGALE parameters following what was used in \citet{Yang2022} when modeling mid-IR detected type 2 AGN. Most of the parameters were chosen to be the default values to reduce the degeneracy besides the parameters discussed below. 

When modeling the stellar emission, we selected 1 Gyr and 2 Gyr for the age of the main stellar population in the galaxy as the age of J1714+6027 is 2.1~Gyr at $z$ = 2.99. We found a huge stellar population metalicity of $Z_{\rm gas}$ = 0.051 (the maximum allowed value in CIGALE), which is in agreement with the best-fit Stellar Age $t$ = 2~Gyr, suggesting a mutual galaxy. 

When modeling the dust emission, we select the {\tt dl2014} model \citep{Draine2014} rather than the default {\tt dale2014} model \citep{Dale2014}. The {\tt dl2014} model has a more complex dust emission model \citep[see discussion in ][]{Boquien2019} and thus has more flexibility in the allowed parameter space compared with {\tt dale2014} model. We found that the {\tt dale2014} model cannot well fit the overall spectral shape of J1714+6027 with a best-fit reduced $\chi^2$ of 2.9. { We note that the property of the dust of the host galaxy is loosely constrained due to the lack of data (and measurements) at rest-frame mid-to-far IR band, where the dust component start to dominate.}

\begingroup
\renewcommand*{\arraystretch}{1.1}
\begin{table*}[htb!]
\centering
\caption{{Fitting parameters values in CIGALE., The best-fit values are labeled in bold.}}
\label{Table:CIGALE}
  \begin{tabular}{lll}
       \hline
       \hline       
      	Module&Parameter&Values\\
       \hline
	SFH&$e$-folding time, $\tau$ (Gyr)&1,{\bf 2},3\\
	&Stellar Age, $t$ (Gyr)&1, {\bf 2}\\
	\hline
	SSP&initial mass function (IMF)&\citet{Chabrier2003}\\
	&Metallicity, $Z_{\rm gas}$&0.02, {\bf 0.05}\\
	\hline
	GDA&Young population &0.1, {\bf 0.2}, 0.3\\
	&powerlaw slope&0, {\bf --0.2}, --0.4\\
	\hline
	Nebula&Ionisation parameter (logU)&--3\\
	&Gas metallicity ($Z_{\rm gas}$)&0.011\\
	\hline
	Dust&Minimum radiation field (Umin)&1, 10, 30, {\bf 50}\\
	&Powerlaw slope, $\alpha$&1, {\bf 2}, 3\\
	&Fraction illuminated&0.5, 0.7, {\bf 0.9}\\
	\hline
	AGN&Optical depth at 9.7 micron&3, {\bf 7}\\
	&Inclination angle, $i$&{\bf 20$^\circ$}, 90$^\circ$\\
	&Angle between the equatorial plane and the edge of the torus, $oa$&40$^\circ$, {\bf 80$^\circ$}\\
	&Disk type&\citet{Schartmann2005}\\
	&Polar direction $E(B-V)$&0.09, {\bf 0.4}\\
	&Polar dust temperature (K)&100, {\bf 1000}, 5000\\
	&Disk slop modification, $\delta$&0.36\\
	&AGN fraction (frac$_{\rm AGN}$)&0.5, {\bf 0.7}, 0.9\\
	\hline
	X-ray&Photon index&1.6\\
	&$\alpha_{\rm OX}$&--1.5, {\bf --1.6}, --1.7\\
	&Maximum allowed deviation of $\alpha_{\rm OX}$ from the empirical relation&0.2\\
	\hline
	Radio&Radio-loudness&5, {\bf 10}, 15\\
	&AGN radio slope&0.6\\
       \hline
	\hline
\end{tabular}
%\tablecomments{The unmentioned parameters of each module are set to the default values in CIGALE.}
\end{table*}
\endgroup

The best-fit results suggest that the AGN is observed through a face-on direction and possess a small covering factor torus as suggested in the X-ray spectral analysis of J1714+6027.

As CIGALE uses a power-law model when fitting the X-ray spectra and requested the flux to be absorption-corrected, we used the fluxes of the intrinsic powerlaw derived from the {\tt borus} model as the input data to CIGALE. When modeling the X-ray component, which should be dominated by the AGN rather than the X-ray binaries in the host galaxy, we fixed the photon index of the source at $\Gamma$ = 1.60 as we derived from the X-ray spectral fitting. We allow the $\alpha_{\rm OX}$ parameter to be free to vary. The radio emission is also dominated by the AGN. We allow the optical radio loudness (RL$_{\rm 30deg}$ = $L_{\nu\rm,5GHz}$/$L_{\nu\rm,2500\AA}$, where $L_{\nu\rm,2500\AA}$ is the AGN 2500~\AA\ intrinsic disk luminosity measured at the inclination angle $i$ = 30$^{\circ}$) free to vary when fitting. 

The dust mass of the host galaxy is not constrained. The best-fit radio loudness is RL$_{\rm 30deg}$ = 10, suggesting that the source is radio quiet.
%mildly radio loud and the source X-ray emission might (partly) be contributed by the jet emission (which is in agreement with the measured very hard X-ray spectral slope). However, we note that the rest-frame UV band (or optical band in the observed frame) of J1714+6027 is dominated by stellar emission, thus the AGN disk luminosity used above to derive the radio loudness is at large uncertainty. The X-ray radio loudness can be defined as $R_X$ = log($L_R$/$L_X$), where $L_R$ is the rest-frame 1.4~GHz luminosity and $L_X$ is the 2--10~keV intrinsic luminosity \citep{DAmato2022}. Sources with $R_X>$ --3.5 are supposed to be radio loud \citep{Terashima2003,Lambrides2020} and the X-ray radio loudness of J1714+6027 is $R_X$ = --3.3, suggesting that the source is mildly radio loud as well. However, we mention that the radio flux of J1714+6027 was measured before 2007 and the source X-ray luminosity in 2011 was much larger than that in 2021. The X-ray radio loudness is $R_X\sim$--4.0 if we adopt the 2011 X-ray luminosity. Therefore, the intrinsic radio loudness of J1714+6027 is uncertain, and future simultaneous X-ray and radio measurement of the source is needed to constrain its radio loudness. 

%%%%%%%%%%%%%%%%%%
\bibliography{referencezxr}{}

\begin{thebibliography}{}
\expandafter\ifx\csname natexlab\endcsname\relax\def\natexlab#1{#1}\fi
\providecommand{\url}[1]{\href{#1}{#1}}
\providecommand{\dodoi}[1]{doi:~\href{http://doi.org/#1}{\nolinkurl{#1}}}
\providecommand{\doeprint}[1]{\href{http://ascl.net/#1}{\nolinkurl{http://ascl.net/#1}}}
\providecommand{\doarXiv}[1]{\href{https://arxiv.org/abs/#1}{\nolinkurl{https://arxiv.org/abs/#1}}}

\bibitem[{Anders \& Grevesse(1989)}]{Anders1989}
Anders, E., \& Grevesse, N. 1989, Geochimica et Cosmochimica Acta, 53, 197 ,
  \dodoi{https://doi.org/10.1016/0016-7037(89)90286-X}

\bibitem[{Annuar {et~al.}(2015)Annuar, Gandhi, Alexander, Lansbury,
  Ar{\'e}valo, Ballantyne, Balokovi{\'c}, Bauer, Boggs, Brandt, Brightman,
  Christensen, Craig, Moro, Hailey, Harrison, Hickox, Matt, Puccetti, Ricci,
  Rigby, Stern, Walton, Zappacosta, \& Zhang}]{Annuar15}
Annuar, A., Gandhi, P., Alexander, D.~M., {et~al.} 2015, The Astrophysical
  Journal, 815, 36.
\newblock \url{http://stacks.iop.org/0004-637X/815/i=1/a=36}

\bibitem[{{Arnaud}(1996)}]{Arnaud1996}
{Arnaud}, K.~A. 1996, Astronomical Data Analysis Software and Systems V, 101,
  17

\bibitem[{{Arrabal Haro} {et~al.}(2023){Arrabal Haro}, {Dickinson},
  {Finkelstein}, {Kartaltepe}, {Donnan}, {Burgarella}, {Carnall}, {Cullen},
  {Dunlop}, {Fern{\'a}ndez}, {Fujimoto}, {Jung}, {Krips}, {Larson}, {Papovich},
  {P{\'e}rez-Gonz{\'a}lez}, {Amor{\'\i}n}, {Bagley}, {Buat}, {Casey},
  {Chworowsky}, {Cohen}, {Ferguson}, {Giavalisco}, {Huertas-Company},
  {Hutchison}, {Kocevski}, {Koekemoer}, {Lucas}, {McLeod}, {McLure}, {Pirzkal},
  {Seill{\'e}}, {Trump}, {Weiner}, {Wilkins}, \& {Zavala}}]{Haro2023}
{Arrabal Haro}, P., {Dickinson}, M., {Finkelstein}, S.~L., {et~al.} 2023, \nat,
  622, 707, \dodoi{10.1038/s41586-023-06521-7}

\bibitem[{Asmus {et~al.}(2015)Asmus, Gandhi, H{\"o}nig, Smette, \&
  Duschl}]{Asmus15}
Asmus, D., Gandhi, P., H{\"o}nig, S.~F., Smette, A., \& Duschl, W.~J. 2015,
  Monthly Notices of the Royal Astronomical Society, 454, 766,
  \dodoi{10.1093/mnras/stv1950}

\bibitem[{{Assef} {et~al.}(2016){Assef}, {Walton}, {Brightman}, {Stern},
  {Alexander}, {Bauer}, {Blain}, {Diaz-Santos}, {Eisenhardt}, {Finkelstein},
  {Hickox}, {Tsai}, \& {Wu}}]{Assef2016}
{Assef}, R.~J., {Walton}, D.~J., {Brightman}, M., {et~al.} 2016, \apj, 819,
  111, \dodoi{10.3847/0004-637X/819/2/111}

\bibitem[{{Assef} {et~al.}(2022){Assef}, {Bauer}, {Blain}, {Brightman},
  {D{\'\i}az-Santos}, {Eisenhardt}, {Jun}, {Stern}, {Tsai}, {Walton}, \&
  {Wu}}]{Assef2022}
{Assef}, R.~J., {Bauer}, F.~E., {Blain}, A.~W., {et~al.} 2022, \apj, 934, 101,
  \dodoi{10.3847/1538-4357/ac77fc}

\bibitem[{Balokovi{\'c} {et~al.}(2018)Balokovi{\'c}, Brightman, Harrison,
  Comastri, Ricci, Buchner, Gandhi, Farrah, \& Stern}]{Borus}
Balokovi{\'c}, M., Brightman, M., Harrison, F.~A., {et~al.} 2018, The
  Astrophysical Journal, 854, 42.
\newblock \url{http://stacks.iop.org/0004-637X/854/i=1/a=42}

\bibitem[{{Boorman} {et~al.}(2023){Boorman}, {Torres-Alb{\`a}}, {Annuar},
  {Marchesi}, {Pfeifle}, {Stern}, {Civano}, {Balokovi{\'c}}, {Buchner},
  {Ricci}, {Alexander}, {Brandt}, {Brightman}, {Chen}, {Creech}, {Gandhi},
  {Garc{\'\i}a}, {Harrison}, {Hickox}, {Kammoun}, {LaMassa}, {Lanzuisi},
  {Marcotulli}, {Madsen}, {Matt}, {Matzeu}, {Nardini}, {Piotrowska},
  {Pizzetti}, {Puccetti}, {Sicilian}, {Silver}, {Walton}, {Wilkins}, \&
  {Zhao}}]{Boorman2023}
{Boorman}, P.~G., {Torres-Alb{\`a}}, N., {Annuar}, A., {et~al.} 2023, arXiv
  e-prints, arXiv:2311.04949, \dodoi{10.48550/arXiv.2311.04949}

\bibitem[{{Boquien} {et~al.}(2019){Boquien}, {Burgarella}, {Roehlly}, {Buat},
  {Ciesla}, {Corre}, {Inoue}, \& {Salas}}]{Boquien2019}
{Boquien}, M., {Burgarella}, D., {Roehlly}, Y., {et~al.} 2019, \aap, 622, A103,
  \dodoi{10.1051/0004-6361/201834156}

\bibitem[{{Bruzual} \& {Charlot}(2003)}]{Bruzual2003}
{Bruzual}, G., \& {Charlot}, S. 2003, \mnras, 344, 1000,
  \dodoi{10.1046/j.1365-8711.2003.06897.x}

\bibitem[{Buchner {et~al.}(2015)Buchner, Georgakakis, Nandra, Brightman,
  Menzel, Liu, Hsu, Salvato, Rangel, Aird, Merloni, \& Ross}]{Buchner2015}
Buchner, J., Georgakakis, A., Nandra, K., {et~al.} 2015, The Astrophysical
  Journal, 802, 89.
\newblock \url{http://stacks.iop.org/0004-637X/802/i=2/a=89}

\bibitem[{{Burgarella} {et~al.}(2005){Burgarella}, {Buat}, \&
  {Iglesias-P{\'a}ramo}}]{Burgarella2005}
{Burgarella}, D., {Buat}, V., \& {Iglesias-P{\'a}ramo}, J. 2005, \mnras, 360,
  1413, \dodoi{10.1111/j.1365-2966.2005.09131.x}

\bibitem[{{Calzetti} {et~al.}(2000){Calzetti}, {Armus}, {Bohlin}, {Kinney},
  {Koornneef}, \& {Storchi-Bergmann}}]{Calzetti2000}
{Calzetti}, D., {Armus}, L., {Bohlin}, R.~C., {et~al.} 2000, \apj, 533, 682,
  \dodoi{10.1086/308692}

\bibitem[{{Casey} {et~al.}(2014){Casey}, {Narayanan}, \& {Cooray}}]{Casey2014}
{Casey}, C.~M., {Narayanan}, D., \& {Cooray}, A. 2014, \physrep, 541, 45,
  \dodoi{10.1016/j.physrep.2014.02.009}

\bibitem[{{Cash}(1979)}]{Cash1979}
{Cash}, W. 1979, \apj, 228, 939

\bibitem[{{Chabrier}(2003)}]{Chabrier2003}
{Chabrier}, G. 2003, \apjl, 586, L133, \dodoi{10.1086/374879}

\bibitem[{{Condon} {et~al.}(2003){Condon}, {Cotton}, {Yin}, {Shupe},
  {Storrie-Lombardi}, {Helou}, {Soifer}, \& {Werner}}]{Condon2003}
{Condon}, J.~J., {Cotton}, W.~D., {Yin}, Q.~F., {et~al.} 2003, \aj, 125, 2411,
  \dodoi{10.1086/374633}

\bibitem[{{Cutri} {et~al.}(2021){Cutri}, {Wright}, {Conrow}, {Fowler},
  {Eisenhardt}, {Grillmair}, {Kirkpatrick}, {Masci}, {McCallon}, {Wheelock},
  {Fajardo-Acosta}, {Yan}, {Benford}, {Harbut}, {Jarrett}, {Lake}, {Leisawitz},
  {Ressler}, {Stanford}, {Tsai}, {Liu}, {Helou}, {Mainzer}, {Gettngs},
  {Gonzalez}, {Hoffman}, {Marsh}, {Padgett}, {Skrutskie}, {Beck}, {Papin}, \&
  {Wittman}}]{Cutri2021}
{Cutri}, R.~M., {Wright}, E.~L., {Conrow}, T., {et~al.} 2021, {VizieR Online
  Data Catalog: AllWISE Data Release (Cutri+ 2013)}, VizieR On-line Data
  Catalog: II/328. Originally published in: IPAC/Caltech (2013)

\bibitem[{{Daddi} {et~al.}(2007){Daddi}, {Alexander}, {Dickinson}, {Gilli},
  {Renzini}, {Elbaz}, {Cimatti}, {Chary}, {Frayer}, {Bauer}, {Brandt},
  {Giavalisco}, {Grogin}, {Huynh}, {Kurk}, {Mignoli}, {Morrison}, {Pope}, \&
  {Ravindranath}}]{Daddi2007}
{Daddi}, E., {Alexander}, D.~M., {Dickinson}, M., {et~al.} 2007, \apj, 670,
  173, \dodoi{10.1086/521820}

\bibitem[{{Dale} {et~al.}(2014){Dale}, {Helou}, {Magdis}, {Armus},
  {D{\'\i}az-Santos}, \& {Shi}}]{Dale2014}
{Dale}, D.~A., {Helou}, G., {Magdis}, G.~E., {et~al.} 2014, \apj, 784, 83,
  \dodoi{10.1088/0004-637X/784/1/83}

\bibitem[{{D'Amato} {et~al.}(2022){D'Amato}, {Prandoni}, {Gilli}, {Vignali},
  {Massardi}, {Liuzzo}, {Jagannathan}, {Brienza}, {Paladino}, {Mignoli},
  {Marchesi}, {Peca}, {Chiaberge}, {Mazzolari}, \& {Norman}}]{DAmato2022}
{D'Amato}, Q., {Prandoni}, I., {Gilli}, R., {et~al.} 2022, \aap, 668, A133,
  \dodoi{10.1051/0004-6361/202244452}

\bibitem[{{Delvecchio} {et~al.}(2014){Delvecchio}, {Gruppioni}, {Pozzi},
  {Berta}, {Zamorani}, {Cimatti}, {Lutz}, {Scott}, {Vignali}, {Cresci},
  {Feltre}, {Cooray}, {Vaccari}, {Fritz}, {Le Floc'h}, {Magnelli}, {Popesso},
  {Oliver}, {Bock}, {Carollo}, {Contini}, {Le F{\'e}vre}, {Lilly}, {Mainieri},
  {Renzini}, \& {Scodeggio}}]{Delvecchio2014}
{Delvecchio}, I., {Gruppioni}, C., {Pozzi}, F., {et~al.} 2014, \mnras, 439,
  2736, \dodoi{10.1093/mnras/stu130}

\bibitem[{{Dey} {et~al.}(2008){Dey}, {Soifer}, {Desai}, {Brand}, {Le Floc'h},
  {Brown}, {Jannuzi}, {Armus}, {Bussmann}, {Brodwin}, {Bian}, {Eisenhardt},
  {Higdon}, {Weedman}, \& {Willner}}]{Dey2008}
{Dey}, A., {Soifer}, B.~T., {Desai}, V., {et~al.} 2008, \apj, 677, 943,
  \dodoi{10.1086/529516}

\bibitem[{{Dey} {et~al.}(2019){Dey}, {Schlegel}, {Lang}, {Blum}, {Burleigh},
  {Fan}, {Findlay}, {Finkbeiner}, {Herrera}, {Juneau}, {Landriau}, {Levi},
  {McGreer}, {Meisner}, {Myers}, {Moustakas}, {Nugent}, {Patej}, {Schlafly},
  {Walker}, {Valdes}, {Weaver}, {Y{\`e}che}, {Zou}, {Zhou}, {Abareshi},
  {Abbott}, {Abolfathi}, {Aguilera}, {Alam}, {Allen}, {Alvarez}, {Annis},
  {Ansarinejad}, {Aubert}, {Beechert}, {Bell}, {BenZvi}, {Beutler}, {Bielby},
  {Bolton}, {Brice{\~n}o}, {Buckley-Geer}, {Butler}, {Calamida}, {Carlberg},
  {Carter}, {Casas}, {Castander}, {Choi}, {Comparat}, {Cukanovaite}, {Delubac},
  {DeVries}, {Dey}, {Dhungana}, {Dickinson}, {Ding}, {Donaldson}, {Duan},
  {Duckworth}, {Eftekharzadeh}, {Eisenstein}, {Etourneau}, {Fagrelius},
  {Farihi}, {Fitzpatrick}, {Font-Ribera}, {Fulmer}, {G{\"a}nsicke},
  {Gaztanaga}, {George}, {Gerdes}, {Gontcho}, {Gorgoni}, {Green}, {Guy},
  {Harmer}, {Hernandez}, {Honscheid}, {Huang}, {James}, {Jannuzi}, {Jiang},
  {Joyce}, {Karcher}, {Karkar}, {Kehoe}, {Kneib}, {Kueter-Young}, {Lan},
  {Lauer}, {Le Guillou}, {Le Van Suu}, {Lee}, {Lesser}, {Perreault Levasseur},
  {Li}, {Mann}, {Marshall}, {Mart{\'\i}nez-V{\'a}zquez}, {Martini}, {du Mas des
  Bourboux}, {McManus}, {Meier}, {M{\'e}nard}, {Metcalfe},
  {Mu{\~n}oz-Guti{\'e}rrez}, {Najita}, {Napier}, {Narayan}, {Newman}, {Nie},
  {Nord}, {Norman}, {Olsen}, {Paat}, {Palanque-Delabrouille}, {Peng},
  {Poppett}, {Poremba}, {Prakash}, {Rabinowitz}, {Raichoor}, {Rezaie},
  {Robertson}, {Roe}, {Ross}, {Ross}, {Rudnick}, {Safonova}, {Saha},
  {S{\'a}nchez}, {Savary}, {Schweiker}, {Scott}, {Seo}, {Shan}, {Silva},
  {Slepian}, {Soto}, {Sprayberry}, {Staten}, {Stillman}, {Stupak}, {Summers},
  {Sien Tie}, {Tirado}, {Vargas-Maga{\~n}a}, {Vivas}, {Wechsler}, {Williams},
  {Yang}, {Yang}, {Yapici}, {Zaritsky}, {Zenteno}, {Zhang}, {Zhang}, {Zhou}, \&
  {Zhou}}]{Dey2019}
{Dey}, A., {Schlegel}, D.~J., {Lang}, D., {et~al.} 2019, \aj, 157, 168,
  \dodoi{10.3847/1538-3881/ab089d}

\bibitem[{{Di Matteo} {et~al.}(2005){Di Matteo}, {Springel}, \&
  {Hernquist}}]{Di_Matteo2005}
{Di Matteo}, T., {Springel}, V., \& {Hernquist}, L. 2005, \nat, 433, 604,
  \dodoi{10.1038/nature03335}

\bibitem[{{Draine} {et~al.}(2007){Draine}, {Dale}, {Bendo}, {Gordon}, {Smith},
  {Armus}, {Engelbracht}, {Helou}, {Kennicutt}, {Li}, {Roussel}, {Walter},
  {Calzetti}, {Moustakas}, {Murphy}, {Rieke}, {Bot}, {Hollenbach}, {Sheth}, \&
  {Teplitz}}]{Draine2007}
{Draine}, B.~T., {Dale}, D.~A., {Bendo}, G., {et~al.} 2007, \apj, 663, 866,
  \dodoi{10.1086/518306}

\bibitem[{{Draine} {et~al.}(2014){Draine}, {Aniano}, {Krause}, {Groves},
  {Sandstrom}, {Braun}, {Leroy}, {Klaas}, {Linz}, {Rix}, {Schinnerer},
  {Schmiedeke}, \& {Walter}}]{Draine2014}
{Draine}, B.~T., {Aniano}, G., {Krause}, O., {et~al.} 2014, \apj, 780, 172,
  \dodoi{10.1088/0004-637X/780/2/172}

\bibitem[{{Duras} {et~al.}(2020){Duras}, {Bongiorno}, {Ricci}, {Piconcelli},
  {Shankar}, {Lusso}, {Bianchi}, {Fiore}, {Maiolino}, {Marconi}, {Onori},
  {Sani}, {Schneider}, {Vignali}, \& {La Franca}}]{Duras2020}
{Duras}, F., {Bongiorno}, A., {Ricci}, F., {et~al.} 2020, \aap, 636, A73,
  \dodoi{10.1051/0004-6361/201936817}

\bibitem[{{Eisenhardt} {et~al.}(2012){Eisenhardt}, {Wu}, {Tsai}, {Assef},
  {Benford}, {Blain}, {Bridge}, {Condon}, {Cushing}, {Cutri}, {Evans},
  {Gelino}, {Griffith}, {Grillmair}, {Jarrett}, {Lonsdale}, {Masci}, {Mason},
  {Petty}, {Sayers}, {Stanford}, {Stern}, {Wright}, \& {Yan}}]{Eisenhardt2012}
{Eisenhardt}, P. R.~M., {Wu}, J., {Tsai}, C.-W., {et~al.} 2012, \apj, 755, 173,
  \dodoi{10.1088/0004-637X/755/2/173}

\bibitem[{{Fabian}(2012)}]{Fabian2012}
{Fabian}, A.~C. 2012, \araa, 50, 455,
  \dodoi{10.1146/annurev-astro-081811-125521}

\bibitem[{{Fabian} {et~al.}(2009){Fabian}, {Zoghbi}, {Ross}, {Uttley}, {Gallo},
  {Brandt}, {Blustin}, {Boller}, {Caballero-Garcia}, {Larsson}, {Miller},
  {Miniutti}, {Ponti}, {Reis}, {Reynolds}, {Tanaka}, \& {Young}}]{Fabian2009}
{Fabian}, A.~C., {Zoghbi}, A., {Ross}, R.~R., {et~al.} 2009, \nat, 459, 540,
  \dodoi{10.1038/nature08007}

\bibitem[{{Fadda} {et~al.}(2006){Fadda}, {Marleau}, {Storrie-Lombardi},
  {Makovoz}, {Frayer}, {Appleton}, {Armus}, {Chapman}, {Choi}, {Fang},
  {Heinrichsen}, {Helou}, {Im}, {Lacy}, {Shupe}, {Soifer}, {Squires}, {Surace},
  {Teplitz}, {Wilson}, \& {Yan}}]{Fadda2006}
{Fadda}, D., {Marleau}, F.~R., {Storrie-Lombardi}, L.~J., {et~al.} 2006, \aj,
  131, 2859, \dodoi{10.1086/504034}

\bibitem[{{Fiore} {et~al.}(2009){Fiore}, {Puccetti}, {Brusa}, {Salvato},
  {Zamorani}, {Aldcroft}, {Aussel}, {Brunner}, {Capak}, {Cappelluti}, {Civano},
  {Comastri}, {Elvis}, {Feruglio}, {Finoguenov}, {Fruscione}, {Gilli},
  {Hasinger}, {Koekemoer}, {Kartaltepe}, {Ilbert}, {Impey}, {Le Floc'h},
  {Lilly}, {Mainieri}, {Martinez-Sansigre}, {McCracken}, {Menci}, {Merloni},
  {Miyaji}, {Sanders}, {Sargent}, {Schinnerer}, {Scoville}, {Silverman},
  {Smolcic}, {Steffen}, {Santini}, {Taniguchi}, {Thompson}, {Trump}, {Vignali},
  {Urry}, \& {Yan}}]{Fiore2009}
{Fiore}, F., {Puccetti}, S., {Brusa}, M., {et~al.} 2009, \apj, 693, 447,
  \dodoi{10.1088/0004-637X/693/1/447}

\bibitem[{{Florez} {et~al.}(2020){Florez}, {Jogee}, {Sherman}, {Stevans},
  {Finkelstein}, {Papovich}, {Kawinwanichakij}, {Ciardullo}, {Gronwall},
  {Urry}, {Kirkpatrick}, {LaMassa}, {Ananna}, \& {Wold}}]{Florez2020}
{Florez}, J., {Jogee}, S., {Sherman}, S., {et~al.} 2020, \mnras, 497, 3273,
  \dodoi{10.1093/mnras/staa2200}

\bibitem[{{Frayer} {et~al.}(2006){Frayer}, {Fadda}, {Yan}, {Marleau}, {Choi},
  {Helou}, {Soifer}, {Appleton}, {Armus}, {Beck}, {Dole}, {Engelbracht},
  {Fang}, {Gordon}, {Heinrichsen}, {Henderson}, {Hesselroth}, {Im}, {Kelly},
  {Lacy}, {Laine}, {Latter}, {Mahoney}, {Makovoz}, {Masci}, {Morrison},
  {Moshir}, {Noriega-Crespo}, {Padgett}, {Pesenson}, {Shupe}, {Squires},
  {Storrie-Lombardi}, {Surace}, {Teplitz}, \& {Wilson}}]{Frayer2006}
{Frayer}, D.~T., {Fadda}, D., {Yan}, L., {et~al.} 2006, \aj, 131, 250,
  \dodoi{10.1086/498690}

\bibitem[{{Gandhi} {et~al.}(2009){Gandhi}, {Horst, H.}, {Smette, A.}, {H\"onig,
  S.}, {Comastri, A.}, {Gilli, R.}, {Vignali, C.}, \& {Duschl, W.}}]{Gandhi09}
{Gandhi}, P., {Horst, H.}, {Smette, A.}, {et~al.} 2009, A\&A, 502, 457,
  \dodoi{10.1051/0004-6361/200811368}

\bibitem[{{Garn} {et~al.}(2007){Garn}, {Green}, {Hales}, {Riley}, \&
  {Alexander}}]{Garn2007}
{Garn}, T., {Green}, D.~A., {Hales}, S. E.~G., {Riley}, J.~M., \& {Alexander},
  P. 2007, \mnras, 376, 1251, \dodoi{10.1111/j.1365-2966.2007.11514.x}

\bibitem[{{Gilli} {et~al.}(2007){Gilli}, {Comastri, A.}, \& {Hasinger,
  G.}}]{gilli07}
{Gilli}, R., {Comastri, A.}, \& {Hasinger, G.} 2007, A\&A, 463, 79,
  \dodoi{10.1051/0004-6361:20066334}

\bibitem[{{Gilli} {et~al.}(2011){Gilli}, {Su}, {Norman}, {Vignali}, {Comastri},
  {Tozzi}, {Rosati}, {Stiavelli}, {Brandt}, {Xue}, {Luo}, {Castellano},
  {Fontana}, {Fiore}, {Mainieri}, \& {Ptak}}]{Gilli2011}
{Gilli}, R., {Su}, J., {Norman}, C., {et~al.} 2011, \apjl, 730, L28,
  \dodoi{10.1088/2041-8205/730/2/L28}

\bibitem[{{Gilli} {et~al.}(2022){Gilli}, {Norman}, {Calura}, {Vito}, {Decarli},
  {Marchesi}, {Iwasawa}, {Comastri}, {Lanzuisi}, {Pozzi}, {D'Amato}, {Vignali},
  {Brusa}, {Mignoli}, \& {Cox}}]{Gilli2022}
{Gilli}, R., {Norman}, C., {Calura}, F., {et~al.} 2022, \aap, 666, A17,
  \dodoi{10.1051/0004-6361/202243708}

\bibitem[{{Granato} {et~al.}(2004){Granato}, {De Zotti}, {Silva}, {Bressan}, \&
  {Danese}}]{Granato2004}
{Granato}, G.~L., {De Zotti}, G., {Silva}, L., {Bressan}, A., \& {Danese}, L.
  2004, \apj, 600, 580, \dodoi{10.1086/379875}

\bibitem[{{HI4PI Collaboration} {et~al.}(2016){HI4PI Collaboration}, {Ben
  Bekhti, N.}, {Fl\"oer, L.}, {Keller, R.}, {Kerp, J.}, {Lenz, D.}, {Winkel,
  B.}, {Bailin, J.}, {Calabretta, M. R.}, {Dedes, L.}, {Ford, H. A.}, {Gibson,
  B. K.}, {Haud, U.}, {Janowiecki, S.}, {Kalberla, P. M. W.}, {Lockman, F. J.},
  {McClure-Griffiths, N. M.}, {Murphy, T.}, {Nakanishi, H.}, {Pisano, D. J.},
  \& {Staveley-Smith, L.}}]{nh}
{HI4PI Collaboration}, {Ben Bekhti, N.}, {Fl\"oer, L.}, {et~al.} 2016, A\&A,
  594, A116, \dodoi{10.1051/0004-6361/201629178}

\bibitem[{Hickox \& Alexander(2018)}]{Hickox18}
Hickox, R.~C., \& Alexander, D.~M. 2018, Annual Review of Astronomy and
  Astrophysics, 56, 625, \dodoi{10.1146/annurev-astro-081817-051803}

\bibitem[{{Inoue}(2011)}]{Inoue2011}
{Inoue}, A.~K. 2011, \mnras, 415, 2920,
  \dodoi{10.1111/j.1365-2966.2011.18906.x}

\bibitem[{{Jansen} {et~al.}(2001){Jansen}, {Lumb, D.}, {Altieri, B.}, {Clavel,
  J.}, {Ehle, M.}, {Erd, C.}, {Gabriel, C.}, {Guainazzi, M.}, {Gondoin, P.},
  {Much, R.}, {Munoz, R.}, {Santos, M.}, {Schartel, N.}, {Texier, D.}, \&
  {Vacanti, G.}}]{SAS}
{Jansen}, F., {Lumb, D.}, {Altieri, B.}, {et~al.} 2001, A\&A, 365, L1,
  \dodoi{10.1051/0004-6361:20000036}

\bibitem[{{Just} {et~al.}(2007){Just}, {Brandt}, {Shemmer}, {Steffen},
  {Schneider}, {Chartas}, \& {Garmire}}]{Just2007}
{Just}, D.~W., {Brandt}, W.~N., {Shemmer}, O., {et~al.} 2007, \apj, 665, 1004,
  \dodoi{10.1086/519990}

\bibitem[{{Lacy} {et~al.}(2011){Lacy}, {Petric}, {Mart{\'\i}nez-Sansigre},
  {Ridgway}, {Sajina}, {Urrutia}, \& {Farrah}}]{Lacy2011}
{Lacy}, M., {Petric}, A.~O., {Mart{\'\i}nez-Sansigre}, A., {et~al.} 2011, \aj,
  142, 196, \dodoi{10.1088/0004-6256/142/6/196}

\bibitem[{{Lacy} {et~al.}(2007){Lacy}, {Petric}, {Sajina}, {Canalizo},
  {Storrie-Lombardi}, {Armus}, {Fadda}, \& {Marleau}}]{Lacy2007}
{Lacy}, M., {Petric}, A.~O., {Sajina}, A., {et~al.} 2007, \aj, 133, 186,
  \dodoi{10.1086/509617}

\bibitem[{{Lacy} {et~al.}(2005){Lacy}, {Wilson}, {Masci}, {Storrie-Lombardi},
  {Appleton}, {Armus}, {Chapman}, {Choi}, {Fadda}, {Fang}, {Frayer},
  {Heinrichsen}, {Helou}, {Im}, {Laine}, {Marleau}, {Shupe}, {Soifer},
  {Squires}, {Surace}, {Teplitz}, \& {Yan}}]{Lacy2005}
{Lacy}, M., {Wilson}, G., {Masci}, F., {et~al.} 2005, \apjs, 161, 41,
  \dodoi{10.1086/432894}

\bibitem[{{Lambrides} {et~al.}(2020){Lambrides}, {Chiaberge}, {Heckman},
  {Gilli}, {Vito}, \& {Norman}}]{Lambrides2020}
{Lambrides}, E.~L., {Chiaberge}, M., {Heckman}, T., {et~al.} 2020, \apj, 897,
  160, \dodoi{10.3847/1538-4357/ab919c}

\bibitem[{{Lanzuisi} {et~al.}(2009){Lanzuisi}, {Piconcelli}, {Fiore},
  {Feruglio}, {Vignali}, {Salvato}, \& {Gruppioni}}]{Lanzuisi2009}
{Lanzuisi}, G., {Piconcelli}, E., {Fiore}, F., {et~al.} 2009, \aap, 498, 67,
  \dodoi{10.1051/0004-6361/200811282}

\bibitem[{{Lanzuisi} {et~al.}(2017){Lanzuisi}, {Delvecchio}, {Berta}, {Brusa},
  {Comastri}, {Gilli}, {Gruppioni}, {Marchesi}, {Perna}, {Pozzi}, {Salvato},
  {Symeonidis}, {Vignali}, {Vito}, {Volonteri}, \& {Zamorani}}]{Lanzuisi2017}
{Lanzuisi}, G., {Delvecchio}, I., {Berta}, S., {et~al.} 2017, \aap, 602, A123,
  \dodoi{10.1051/0004-6361/201629955}

\bibitem[{{Lanzuisi} {et~al.}(2018){Lanzuisi}, {Civano}, {Marchesi},
  {Comastri}, {Brusa}, {Gilli}, {Vignali}, {Zamorani}, {Brightman},
  {Griffiths}, \& {Koekemoer}}]{Lanzuisi2018}
{Lanzuisi}, G., {Civano}, F., {Marchesi}, S., {et~al.} 2018, \mnras, 480, 2578,
  \dodoi{10.1093/mnras/sty2025}

\bibitem[{{Lanzuisi} {et~al.}(2019){Lanzuisi}, {Gilli}, {Cappi}, {Dadina},
  {Bianchi}, {Brusa}, {Chartas}, {Civano}, {Comastri}, {Marinucci}, {Middei},
  {Piconcelli}, {Vignali}, {Brandt}, {Tombesi}, \& {Gaspari}}]{Lanzuisi2019}
{Lanzuisi}, G., {Gilli}, R., {Cappi}, M., {et~al.} 2019, \apjl, 875, L20,
  \dodoi{10.3847/2041-8213/ab15dc}

\bibitem[{{Leitherer} {et~al.}(2002){Leitherer}, {Li}, {Calzetti}, \&
  {Heckman}}]{Leitherer2002}
{Leitherer}, C., {Li}, I.~H., {Calzetti}, D., \& {Heckman}, T.~M. 2002, \apjs,
  140, 303, \dodoi{10.1086/342486}

\bibitem[{{Lightman} \& {White}(1988)}]{Lightman1988}
{Lightman}, A.~P., \& {White}, T.~R. 1988, \apj, 335, 57,
  \dodoi{10.1086/166905}

\bibitem[{{Lusso} \& {Risaliti}(2017)}]{Lusso2017}
{Lusso}, E., \& {Risaliti}, G. 2017, \aap, 602, A79,
  \dodoi{10.1051/0004-6361/201630079}

\bibitem[{{Lusso} {et~al.}(2012){Lusso}, {Comastri}, {Simmons}, {Mignoli},
  {Zamorani}, {Vignali}, {Brusa}, {Shankar}, {Lutz}, {Trump}, {Maiolino},
  {Gilli}, {Bolzonella}, {Puccetti}, {Salvato}, {Impey}, {Civano}, {Elvis},
  {Mainieri}, {Silverman}, {Koekemoer}, {Bongiorno}, {Merloni}, {Berta}, {Le
  Floc'h}, {Magnelli}, {Pozzi}, \& {Riguccini}}]{Lusso2012}
{Lusso}, E., {Comastri}, A., {Simmons}, B.~D., {et~al.} 2012, \mnras, 425, 623,
  \dodoi{10.1111/j.1365-2966.2012.21513.x}

\bibitem[{{Lutz} {et~al.}(2004){Lutz}, {Maiolino}, {Spoon}, \&
  {Moorwood}}]{Lutz2004}
{Lutz}, D., {Maiolino}, R., {Spoon}, H.~W.~W., \& {Moorwood}, A.~F.~M. 2004,
  \aap, 418, 465, \dodoi{10.1051/0004-6361:20035838}

\bibitem[{{Madau} \& {Dickinson}(2014)}]{Madau2014}
{Madau}, P., \& {Dickinson}, M. 2014, \araa, 52, 415,
  \dodoi{10.1146/annurev-astro-081811-125615}

\bibitem[{{Madsen} {et~al.}(2017){Madsen}, {Beardmore}, {Forster}, {Guainazzi},
  {Marshall}, {Miller}, {Page}, \& {Stuhlinger}}]{Madsen2017}
{Madsen}, K.~K., {Beardmore}, A.~P., {Forster}, K., {et~al.} 2017, \aj, 153, 2,
  \dodoi{10.3847/1538-3881/153/1/2}

\bibitem[{Magdziarz \& Zdziarski(1995)}]{pexrav}
Magdziarz, P., \& Zdziarski, A.~A. 1995, Monthly Notices of the Royal
  Astronomical Society, 273, 837, \dodoi{10.1093/mnras/273.3.837}

\bibitem[{Marchesi {et~al.}(2018)Marchesi, Ajello, Marcotulli, Comastri,
  Lanzuisi, \& Vignali}]{Marchesi2018}
Marchesi, S., Ajello, M., Marcotulli, L., {et~al.} 2018, The Astrophysical
  Journal, 854, 49.
\newblock \url{http://stacks.iop.org/0004-637X/854/i=1/a=49}

\bibitem[{Marchesi {et~al.}(2019)Marchesi, Ajello, Zhao, Marcotulli,
  Balokovi{\'{c}}, Brightman, Comastri, Cusumano, Lanzuisi, Parola, Segreto, \&
  Vignali}]{Marchesi_2019}
Marchesi, S., Ajello, M., Zhao, X., {et~al.} 2019, The Astrophysical Journal,
  872, 8, \dodoi{10.3847/1538-4357/aafbeb}

\bibitem[{{Masoura} {et~al.}(2021){Masoura}, {Mountrichas}, {Georgantopoulos},
  \& {Plionis}}]{Masoura2021}
{Masoura}, V.~A., {Mountrichas}, G., {Georgantopoulos}, I., \& {Plionis}, M.
  2021, \aap, 646, A167, \dodoi{10.1051/0004-6361/202039238}

\bibitem[{{Menci} {et~al.}(2008){Menci}, {Fiore}, {Puccetti}, \&
  {Cavaliere}}]{Menci2008}
{Menci}, N., {Fiore}, F., {Puccetti}, S., \& {Cavaliere}, A. 2008, \apj, 686,
  219, \dodoi{10.1086/591438}

\bibitem[{{Merloni} \& {Heinz}(2008)}]{Merloni2008}
{Merloni}, A., \& {Heinz}, S. 2008, \mnras, 388, 1011,
  \dodoi{10.1111/j.1365-2966.2008.13472.x}

\bibitem[{{Middei} {et~al.}(2017){Middei}, {Vagnetti}, {Bianchi}, {La Franca},
  {Paolillo}, \& {Ursini}}]{Middei2017}
{Middei}, R., {Vagnetti}, F., {Bianchi}, S., {et~al.} 2017, \aap, 599, A82,
  \dodoi{10.1051/0004-6361/201629940}

\bibitem[{{Mortlock} {et~al.}(2011){Mortlock}, {Warren}, {Venemans}, {Patel},
  {Hewett}, {McMahon}, {Simpson}, {Theuns}, {Gonz{\'a}les-Solares}, {Adamson},
  {Dye}, {Hambly}, {Hirst}, {Irwin}, {Kuiper}, {Lawrence}, \&
  {R{\"o}ttgering}}]{Mortlock2011}
{Mortlock}, D.~J., {Warren}, S.~J., {Venemans}, B.~P., {et~al.} 2011, \nat,
  474, 616, \dodoi{10.1038/nature10159}

\bibitem[{Murphy \& Yaqoob(2009)}]{MYTorus2009}
Murphy, K.~D., \& Yaqoob, T. 2009, Monthly Notices of the Royal Astronomical
  Society, 397, 1549, \dodoi{10.1111/j.1365-2966.2009.15025.x}

\bibitem[{{Noll} {et~al.}(2009){Noll}, {Burgarella}, {Giovannoli}, {Buat},
  {Marcillac}, \& {Mu{\~n}oz-Mateos}}]{Noll2009}
{Noll}, S., {Burgarella}, D., {Giovannoli}, E., {et~al.} 2009, \aap, 507, 1793,
  \dodoi{10.1051/0004-6361/200912497}

\bibitem[{{Paolillo} {et~al.}(2023){Paolillo}, {Papadakis}, {Brandt}, {Bauer},
  {Lanzuisi}, {Allevato}, {Shemmer}, {Zheng}, {De Cicco}, {Gilli}, {Luo},
  {Thomas}, {Tozzi}, {Vito}, \& {Xue}}]{Paolillo2023}
{Paolillo}, M., {Papadakis}, I.~E., {Brandt}, W.~N., {et~al.} 2023, \aap, 673,
  A68, \dodoi{10.1051/0004-6361/202245291}

\bibitem[{{Piconcelli} {et~al.}(2015){Piconcelli}, {Vignali}, {Bianchi},
  {Zappacosta}, {Fritz}, {Lanzuisi}, {Miniutti}, {Bongiorno}, {Feruglio},
  {Fiore}, \& {Maiolino}}]{Piconcelli2015}
{Piconcelli}, E., {Vignali}, C., {Bianchi}, S., {et~al.} 2015, \aap, 574, L9,
  \dodoi{10.1051/0004-6361/201425324}

\bibitem[{{Pouliasis} {et~al.}(2022){Pouliasis}, {Mountrichas},
  {Georgantopoulos}, {Ruiz}, {Gilli}, {Koulouridis}, {Akiyama}, {Ueda},
  {Garrel}, {Nagao}, {Paltani}, {Pierre}, {Toba}, \& {Vignali}}]{Pouliasis2022}
{Pouliasis}, E., {Mountrichas}, G., {Georgantopoulos}, I., {et~al.} 2022, \aap,
  667, A56, \dodoi{10.1051/0004-6361/202243502}

\bibitem[{{Pozzi} {et~al.}(2007){Pozzi}, {Vignali}, {Comastri}, {Pozzetti},
  {Mignoli}, {Gruppioni}, {Zamorani}, {Lari}, {Civano}, {Brusa}, {Fiore},
  {Maiolino}, \& {La Franca}}]{Pozzi2007}
{Pozzi}, F., {Vignali}, C., {Comastri}, A., {et~al.} 2007, \aap, 468, 603,
  \dodoi{10.1051/0004-6361:20077092}

\bibitem[{Puccetti {et~al.}(2014)Puccetti, Comastri, Fiore, Ar{\'e}valo,
  Risaliti, Bauer, Brandt, Stern, Harrison, Alexander, Boggs, Christensen,
  Craig, Gandhi, Hailey, Koss, Lansbury, Luo, Madejski, Matt, Walton, \&
  Zhang}]{puccetti14}
Puccetti, S., Comastri, A., Fiore, F., {et~al.} 2014, The Astrophysical
  Journal, 793, 26.
\newblock \url{http://stacks.iop.org/0004-637X/793/i=1/a=26}

\bibitem[{Ricci {et~al.}(2017{\natexlab{a}})Ricci, Trakhtenbrot, Koss, Ueda,
  Vecchio, Treister, Schawinski, Paltani, Oh, Lamperti, Berney, Gandhi,
  Ichikawa, Bauer, Ho, Asmus, Beckmann, Soldi, Balokovi{\'c}, Gehrels, \&
  Markwardt}]{Ricci2017}
Ricci, C., Trakhtenbrot, B., Koss, M.~J., {et~al.} 2017{\natexlab{a}}, The
  Astrophysical Journal Supplement Series, 233, 17.
\newblock \url{http://stacks.iop.org/0067-0049/233/i=2/a=17}

\bibitem[{Ricci {et~al.}(2017{\natexlab{b}})Ricci, Assef, Stern, Nikutta,
  Alexander, Asmus, Ballantyne, Bauer, Blain, Boggs, Boorman, Brandt,
  Brightman, Chang, Chen, Christensen, Comastri, Craig, D{\'{\i}}az-Santos,
  Eisenhardt, Farrah, Gandhi, Hailey, Harrison, Jun, Koss, LaMassa, Lansbury,
  Markwardt, Stalevski, Stanley, Treister, Tsai, Walton, Wu, Zappacosta, \&
  Zhang}]{Ricci_2017}
Ricci, C., Assef, R.~J., Stern, D., {et~al.} 2017{\natexlab{b}}, The
  Astrophysical Journal, 835, 105, \dodoi{10.3847/1538-4357/835/1/105}

\bibitem[{{Salim} \& {Narayanan}(2020)}]{Salim2020}
{Salim}, S., \& {Narayanan}, D. 2020, \araa, 58, 529,
  \dodoi{10.1146/annurev-astro-032620-021933}

\bibitem[{{Salpeter}(1964)}]{Salpeter1964}
{Salpeter}, E.~E. 1964, \apj, 140, 796, \dodoi{10.1086/147973}

\bibitem[{{Schartmann} {et~al.}(2005){Schartmann}, {Meisenheimer}, {Camenzind},
  {Wolf}, \& {Henning}}]{Schartmann2005}
{Schartmann}, M., {Meisenheimer}, K., {Camenzind}, M., {Wolf}, S., \&
  {Henning}, T. 2005, \aap, 437, 861, \dodoi{10.1051/0004-6361:20042363}

\bibitem[{{Schreiber} {et~al.}(2015){Schreiber}, {Pannella}, {Elbaz},
  {B{\'e}thermin}, {Inami}, {Dickinson}, {Magnelli}, {Wang}, {Aussel}, {Daddi},
  {Juneau}, {Shu}, {Sargent}, {Buat}, {Faber}, {Ferguson}, {Giavalisco},
  {Koekemoer}, {Magdis}, {Morrison}, {Papovich}, {Santini}, \&
  {Scott}}]{Schreiber2015}
{Schreiber}, C., {Pannella}, M., {Elbaz}, D., {et~al.} 2015, \aap, 575, A74,
  \dodoi{10.1051/0004-6361/201425017}

\bibitem[{{Shirley} {et~al.}(2021){Shirley}, {Duncan}, {Campos Varillas},
  {Hurley}, {Ma{\l}ek}, {Roehlly}, {Smith}, {Aussel}, {Bakx}, {Buat},
  {Burgarella}, {Christopher}, {Duivenvoorden}, {Eales}, {Efstathiou},
  {Gonz{\'a}lez Solares}, {Griffin}, {Jarvis}, {Faro}, {Marchetti}, {McCheyne},
  {Papadopoulos}, {Penner}, {Pons}, {Prescott}, {Rigby}, {Rottgering},
  {Saxena}, {Scudder}, {Vaccari}, {Wang}, \& {Oliver}}]{Shirley2021}
{Shirley}, R., {Duncan}, K., {Campos Varillas}, M.~C., {et~al.} 2021, \mnras,
  507, 129, \dodoi{10.1093/mnras/stab1526}

\bibitem[{{Snios} {et~al.}(2020){Snios}, {Siemiginowska}, {Sobolewska},
  {Cheung}, {Kashyap}, {Migliori}, {Schwartz}, {Stawarz}, \&
  {Worrall}}]{Snios2020}
{Snios}, B., {Siemiginowska}, A., {Sobolewska}, M., {et~al.} 2020, \apj, 899,
  127, \dodoi{10.3847/1538-4357/aba2ca}

\bibitem[{{Stalevski} {et~al.}(2012){Stalevski}, {Fritz}, {Baes}, {Nakos}, \&
  {Popovi{\'c}}}]{Stalevski2012}
{Stalevski}, M., {Fritz}, J., {Baes}, M., {Nakos}, T., \& {Popovi{\'c}},
  L.~{\v{C}}. 2012, \mnras, 420, 2756, \dodoi{10.1111/j.1365-2966.2011.19775.x}

\bibitem[{{Stalevski} {et~al.}(2016){Stalevski}, {Ricci}, {Ueda}, {Lira},
  {Fritz}, \& {Baes}}]{Stalevski2016}
{Stalevski}, M., {Ricci}, C., {Ueda}, Y., {et~al.} 2016, \mnras, 458, 2288,
  \dodoi{10.1093/mnras/stw444}

\bibitem[{{Steffen} {et~al.}(2006){Steffen}, {Strateva}, {Brandt}, {Alexander},
  {Koekemoer}, {Lehmer}, {Schneider}, \& {Vignali}}]{Steffen2006}
{Steffen}, A.~T., {Strateva}, I., {Brandt}, W.~N., {et~al.} 2006, \aj, 131,
  2826, \dodoi{10.1086/503627}

\bibitem[{{Stern}(2015)}]{Stern2015}
{Stern}, D. 2015, \apj, 807, 129, \dodoi{10.1088/0004-637X/807/2/129}

\bibitem[{{Str\"uder} {et~al.}(2001){Str\"uder}, {Briel, U.}, {Dennerl, K.},
  {Hartmann, R.}, {Kendziorra, E.}, {Meidinger, N.}, {Pfeffermann, E.},
  {Reppin, C.}, {Aschenbach, B.}, {Bornemann, W.}, {Br\"auninger, H.},
  {Burkert, W.}, {Elender, M.}, {Freyberg, M.}, {Haberl, F.}, {Hartner, G.},
  {Heuschmann, F.}, {Hippmann, H.}, {Kastelic, E.}, {Kemmer, S.}, {Kettenring,
  G.}, {Kink, W.}, {Krause, N.}, {M\"uller, S.}, {Oppitz, A.}, {Pietsch, W.},
  {Popp, M.}, {Predehl, P.}, {Read, A.}, {Stephan, K. H.}, {St\"otter, D.},
  {Tr\"umper, J.}, {Holl, P.}, {Kemmer, J.}, {Soltau, H.}, {St\"otter, R.},
  {Weber, U.}, {Weichert, U.}, {von Zanthier, C.}, {Carathanassis, D.}, {Lutz,
  G.}, {Richter, R. H.}, {Solc, P.}, {B\"ottcher, H.}, {Kuster, M.}, {Staubert,
  R.}, {Abbey, A.}, {Holland, A.}, {Turner, M.}, {Balasini, M.}, {Bignami, G.
  F.}, {La Palombara, N.}, {Villa, G.}, {Buttler, W.}, {Gianini, F.}, {Lain\'e,
  R.}, {Lumb, D.}, \& {Dhez, P.}}]{pn}
{Str\"uder}, L., {Briel, U.}, {Dennerl, K.}, {et~al.} 2001, A\&A, 365, L18,
  \dodoi{10.1051/0004-6361:20000066}

\bibitem[{{Terashima} \& {Wilson}(2003)}]{Terashima2003}
{Terashima}, Y., \& {Wilson}, A.~S. 2003, \apj, 583, 145,
  \dodoi{10.1086/345339}

\bibitem[{{Thorne} {et~al.}(2021){Thorne}, {Robotham}, {Davies}, {Bellstedt},
  {Driver}, {Bravo}, {Bremer}, {Holwerda}, {Hopkins}, {Lagos}, {Phillipps},
  {Siudek}, {Taylor}, \& {Wright}}]{Thorne2021}
{Thorne}, J.~E., {Robotham}, A. S.~G., {Davies}, L. J.~M., {et~al.} 2021,
  \mnras, 505, 540, \dodoi{10.1093/mnras/stab1294}

\bibitem[{{Torres-Alb{\`a}} {et~al.}(2021){Torres-Alb{\`a}}, {Marchesi},
  {Zhao}, {Ajello}, {Silver}, {Ananna}, {Balokovi{\'c}}, {Boorman}, {Comastri},
  {Gilli}, {Lanzuisi}, {Murphy}, {Urry}, \& {Vignali}}]{Nuria}
{Torres-Alb{\`a}}, N., {Marchesi}, S., {Zhao}, X., {et~al.} 2021, arXiv
  e-prints, arXiv:2109.00599.
\newblock \doarXiv{2109.00599}

\bibitem[{{Tsai} {et~al.}(2015){Tsai}, {Eisenhardt}, {Wu}, {Stern}, {Assef},
  {Blain}, {Bridge}, {Benford}, {Cutri}, {Griffith}, {Jarrett}, {Lonsdale},
  {Masci}, {Moustakas}, {Petty}, {Sayers}, {Stanford}, {Wright}, {Yan},
  {Leisawitz}, {Liu}, {Mainzer}, {McLean}, {Padgett}, {Skrutskie}, {Gelino},
  {Beichman}, \& {Juneau}}]{Tsai2015}
{Tsai}, C.-W., {Eisenhardt}, P. R.~M., {Wu}, J., {et~al.} 2015, \apj, 805, 90,
  \dodoi{10.1088/0004-637X/805/2/90}

\bibitem[{{Turner} {et~al.}(2001){Turner}, {Abbey, A.}, {Arnaud, M.},
  {Balasini, M.}, {Barbera, M.}, {Belsole, E.}, {Bennie, P. J.}, {Bernard, J.
  P.}, {Bignami, G. F.}, {Boer, M.}, {Briel, U.}, {Butler, I.}, {Cara, C.},
  {Chabaud, C.}, {Cole, R.}, {Collura, A.}, {Conte, M.}, {Cros, A.}, {Denby,
  M.}, {Dhez, P.}, {Di Coco, G.}, {Dowson, J.}, {Ferrando, P.}, {Ghizzardi,
  S.}, {Gianotti, F.}, {Goodall, C. V.}, {Gretton, L.}, {Griffiths, R. G.},
  {Hainaut, O.}, {Hochedez, J. F.}, {Holland, A. D.}, {Jourdain, E.},
  {Kendziorra, E.}, {Lagostina, A.}, {Laine, R.}, {La Palombara, N.},
  {Lortholary, M.}, {Lumb, D.}, {Marty, P.}, {Molendi, S.}, {Pigot, C.},
  {Poindron, E.}, {Pounds, K. A.}, {Reeves, J. N.}, {Reppin, C.}, {Rothenflug,
  R.}, {Salvetat, P.}, {Sauvageot, J. L.}, {Schmitt, D.}, {Sembay, S.}, {Short,
  A. D. T.}, {Spragg, J.}, {Stephen, J.}, {Str\"uder, L.}, {Tiengo, A.},
  {Trifoglio, M.}, {Tr\"umper, J.}, {Vercellone, S.}, {Vigroux, L.}, {Villa,
  G.}, {Ward, M. J.}, {Whitehead, S.}, \& {Zonca, E.}}]{MOS}
{Turner}, M. J.~L., {Abbey, A.}, {Arnaud, M.}, {et~al.} 2001, A\&A, 365, L27,
  \dodoi{10.1051/0004-6361:20000087}

\bibitem[{{Ulrich} {et~al.}(1997){Ulrich}, {Maraschi}, \& {Urry}}]{Ulrich1997}
{Ulrich}, M.-H., {Maraschi}, L., \& {Urry}, C.~M. 1997, \araa, 35, 445,
  \dodoi{10.1146/annurev.astro.35.1.445}

\bibitem[{Ursini {et~al.}(2018)Ursini, Bassani, Panessa, Bazzano, Bird,
  Malizia, \& Ubertini}]{Ursini18}
Ursini, F., Bassani, L., Panessa, F., {et~al.} 2018, Monthly Notices of the
  Royal Astronomical Society, 474, 5684, \dodoi{10.1093/mnras/stx3159}

\bibitem[{{Vagnetti} {et~al.}(2011){Vagnetti}, {Turriziani}, \&
  {Trevese}}]{Vagnetti2011}
{Vagnetti}, F., {Turriziani}, S., \& {Trevese}, D. 2011, \aap, 536, A84,
  \dodoi{10.1051/0004-6361/201118072}

\bibitem[{Verner {et~al.}(1996)Verner, Ferland, Korista, \&
  Yakovlev}]{Verner1996}
Verner, D., Ferland, G., Korista, K., \& Yakovlev, D. 1996, Astrophysical
  Journal, 465, 487

\bibitem[{{Vito} {et~al.}(2018){Vito}, {Brandt}, {Stern}, {Assef}, {Chen},
  {Brightman}, {Comastri}, {Eisenhardt}, {Garmire}, {Hickox}, {Lansbury},
  {Tsai}, {Walton}, \& {Wu}}]{Vito2018}
{Vito}, F., {Brandt}, W.~N., {Stern}, D., {et~al.} 2018, \mnras, 474, 4528,
  \dodoi{10.1093/mnras/stx3120}

\bibitem[{{Vito} {et~al.}(2020){Vito}, {Brandt}, {Lehmer}, {Vignali}, {Zou},
  {Bauer}, {Bremer}, {Gilli}, {Ivison}, \& {Spingola}}]{Vito2020}
{Vito}, F., {Brandt}, W.~N., {Lehmer}, B.~D., {et~al.} 2020, \aap, 642, A149,
  \dodoi{10.1051/0004-6361/202038848}

\bibitem[{{Wang} {et~al.}(2019){Wang}, {Schreiber}, {Elbaz}, {Yoshimura},
  {Kohno}, {Shu}, {Yamaguchi}, {Pannella}, {Franco}, {Huang}, {Lim}, \&
  {Wang}}]{Wang2019}
{Wang}, T., {Schreiber}, C., {Elbaz}, D., {et~al.} 2019, \nat, 572, 211,
  \dodoi{10.1038/s41586-019-1452-4}

\bibitem[{{Webb} {et~al.}(2020){Webb}, {Coriat}, {Traulsen}, {Ballet}, {Motch},
  {Carrera}, {Koliopanos}, {Authier}, {de la Calle}, {Ceballos}, {Colomo},
  {Chuard}, {Freyberg}, {Garcia}, {Kolehmainen}, {Lamer}, {Lin}, {Maggi},
  {Michel}, {Page}, {Page}, {Perea-Calderon}, {Pineau}, {Rodriguez}, {Rosen},
  {Santos Lleo}, {Saxton}, {Schwope}, {Tom{\'a}s}, {Watson}, \&
  {Zakardjian}}]{Webb2020}
{Webb}, N.~A., {Coriat}, M., {Traulsen}, I., {et~al.} 2020, \aap, 641, A136,
  \dodoi{10.1051/0004-6361/201937353}

\bibitem[{{Werner} {et~al.}(2004){Werner}, {Roellig}, {Low}, {Rieke}, {Rieke},
  {Hoffmann}, {Young}, {Houck}, {Brandl}, {Fazio}, {Hora}, {Gehrz}, {Helou},
  {Soifer}, {Stauffer}, {Keene}, {Eisenhardt}, {Gallagher}, {Gautier}, {Irace},
  {Lawrence}, {Simmons}, {Van Cleve}, {Jura}, {Wright}, \&
  {Cruikshank}}]{Werner2004}
{Werner}, M.~W., {Roellig}, T.~L., {Low}, F.~J., {et~al.} 2004, \apjs, 154, 1,
  \dodoi{10.1086/422992}

\bibitem[{Wik {et~al.}(2014)Wik, Hornstrup, Molendi, Madejski, Harrison,
  Zoglauer, Grefenstette, Gastaldello, Madsen, Westergaard, Ferreira,
  Kitaguchi, Pedersen, Boggs, Christensen, Craig, Hailey, Stern, \&
  Zhang}]{Wik_2014}
Wik, D.~R., Hornstrup, A., Molendi, S., {et~al.} 2014, The Astrophysical
  Journal, 792, 48, \dodoi{10.1088/0004-637x/792/1/48}

\bibitem[{{Wu} {et~al.}(2012){Wu}, {Tsai}, {Sayers}, {Benford}, {Bridge},
  {Blain}, {Eisenhardt}, {Stern}, {Petty}, {Assef}, {Bussmann}, {Comerford},
  {Cutri}, {Evans}, {Griffith}, {Jarrett}, {Lake}, {Lonsdale}, {Rho},
  {Stanford}, {Weiner}, {Wright}, \& {Yan}}]{Wu2012}
{Wu}, J., {Tsai}, C.-W., {Sayers}, J., {et~al.} 2012, \apj, 756, 96,
  \dodoi{10.1088/0004-637X/756/1/96}

\bibitem[{{Yang} {et~al.}(2016){Yang}, {Brandt}, {Luo}, {Xue}, {Bauer}, {Sun},
  {Kim}, {Schulze}, {Zheng}, {Paolillo}, {Shemmer}, {Liu}, {Schneider},
  {Vignali}, {Vito}, \& {Wang}}]{Yang2016}
{Yang}, G., {Brandt}, W.~N., {Luo}, B., {et~al.} 2016, \apj, 831, 145,
  \dodoi{10.3847/0004-637X/831/2/145}

\bibitem[{{Yang} {et~al.}(2020){Yang}, {Boquien}, {Buat}, {Burgarella},
  {Ciesla}, {Duras}, {Stalevski}, {Brandt}, \& {Papovich}}]{Yang2020}
{Yang}, G., {Boquien}, M., {Buat}, V., {et~al.} 2020, \mnras, 491, 740,
  \dodoi{10.1093/mnras/stz3001}

\bibitem[{{Yang} {et~al.}(2022){Yang}, {Boquien}, {Brandt}, {Buat},
  {Burgarella}, {Ciesla}, {Lehmer}, {Ma{\l}ek}, {Mountrichas}, {Papovich},
  {Pons}, {Stalevski}, {Theul{\'e}}, \& {Zhu}}]{Yang2022}
{Yang}, G., {Boquien}, M., {Brandt}, W.~N., {et~al.} 2022, \apj, 927, 192,
  \dodoi{10.3847/1538-4357/ac4971}

\bibitem[{{Yang} {et~al.}(2023){Yang}, {Caputi}, {Papovich}, {Arrabal Haro},
  {Bagley}, {Behroozi}, {Bell}, {Bisigello}, {Buat}, {Burgarella}, {Cheng},
  {Cleri}, {Dav{\'e}}, {Dickinson}, {Elbaz}, {Ferguson}, {Finkelstein},
  {Grogin}, {Hathi}, {Hirschmann}, {Holwerda}, {Huertas-Company}, {Hutchison},
  {Iani}, {Kartaltepe}, {Kirkpatrick}, {Kocevski}, {Koekemoer}, {Kokorev},
  {Larson}, {Lucas}, {P{\'e}rez-Gonz{\'a}lez}, {Rinaldi}, {Shen}, {Trump}, {de
  la Vega}, {Yung}, \& {Zavala}}]{Yang2023}
{Yang}, G., {Caputi}, K.~I., {Papovich}, C., {et~al.} 2023, \apjl, 950, L5,
  \dodoi{10.3847/2041-8213/acd639}

\bibitem[{Yaqoob(2012)}]{MYTorus2012}
Yaqoob, T. 2012, Monthly Notices of the Royal Astronomical Society, 423, 3360,
  \dodoi{10.1111/j.1365-2966.2012.21129.x}

\bibitem[{Zappacosta {et~al.}(2018)Zappacosta, Comastri, Civano, Puccetti,
  Fiore, Aird, Moro, Lansbury, Lanzuisi, Goulding, Mullaney, Stern, Ajello,
  Alexander, Ballantyne, Bauer, Brandt, Chen, Farrah, Harrison, Gandhi, Lanz,
  Masini, Marchesi, Ricci, \& Treister}]{Zappacosta_2018}
Zappacosta, L., Comastri, A., Civano, F., {et~al.} 2018, The Astrophysical
  Journal, 854, 33, \dodoi{10.3847/1538-4357/aaa550}

\bibitem[{{Zappacosta} {et~al.}(2018){Zappacosta}, {Piconcelli}, {Duras},
  {Vignali}, {Valiante}, {Bianchi}, {Bongiorno}, {Fiore}, {Feruglio},
  {Lanzuisi}, {Maiolino}, {Mathur}, {Miniutti}, \& {Ricci}}]{Zappacosta2018}
{Zappacosta}, L., {Piconcelli}, E., {Duras}, F., {et~al.} 2018, \aap, 618, A28,
  \dodoi{10.1051/0004-6361/201732557}

\bibitem[{{Zappacosta} {et~al.}(2020){Zappacosta}, {Piconcelli}, {Giustini},
  {Vietri}, {Duras}, {Miniutti}, {Bischetti}, {Bongiorno}, {Brusa},
  {Chiaberge}, {Comastri}, {Feruglio}, {Luminari}, {Marconi}, {Ricci},
  {Vignali}, \& {Fiore}}]{Zappacosta2020}
{Zappacosta}, L., {Piconcelli}, E., {Giustini}, M., {et~al.} 2020, \aap, 635,
  L5, \dodoi{10.1051/0004-6361/201937292}

\bibitem[{Zhao {et~al.}(2020)Zhao, Marchesi, Ajello, Balokovi{\'c}, \&
  Fischer}]{Zhao_2020}
Zhao, X., Marchesi, S., Ajello, M., Balokovi{\'c}, M., \& Fischer, T. 2020, The
  Astrophysical Journal, 894, 71, \dodoi{10.3847/1538-4357/ab879d}

\bibitem[{{Zhao} {et~al.}(2021){Zhao}, {Marchesi}, {Ajello}, {Cole}, {Hu},
  {Silver}, \& {Torres-Alb{\`a}}}]{Zhao2021a}
{Zhao}, X., {Marchesi}, S., {Ajello}, M., {et~al.} 2021, \aap, 650, A57,
  \dodoi{10.1051/0004-6361/202140297}

\bibitem[{Zhao {et~al.}(2019)Zhao, Marchesi, Ajello, Marcotulli, Cusumano,
  Parola, \& Vignali}]{Zhao_2019_1}
Zhao, X., Marchesi, S., Ajello, M., {et~al.} 2019, The Astrophysical Journal,
  870, 60, \dodoi{10.3847/1538-4357/aaf1a0}

\bibitem[{Zhao {et~al.}(2021)Zhao, Civano, Fornasini, Alexander, Cappelluti,
  Chen, Cohen, Elvis, Gandhi, Grogin, Hickox, Jansen, Koekemoer, Lanzuisi,
  Maksym, Masini, Rosario, Ward, Willmer, \& Windhorst}]{Zhao2021}
Zhao, X., Civano, F., Fornasini, F.~M., {et~al.} 2021, Monthly Notices of the
  Royal Astronomical Society, 508, 5176, \dodoi{10.1093/mnras/stab2885}

\end{thebibliography}
\bibliographystyle{aasjournal}

\end{document}